\documentclass[12pt]{article}
\pdfoutput=1
\usepackage{graphicx,amsmath,amsfonts,amssymb,stmaryrd,latexsym}
\usepackage{url,subcaption,comment}
\usepackage{hyperref}
\usepackage{xcolor}
\hypersetup{
    colorlinks=blue,       
    linkcolor=purple,   
    citecolor=blue,        
    filecolor=magenta,      
    urlcolor=black           
}
\usepackage{cite}
\usepackage{dcolumn}
\usepackage{bm}
\usepackage{booktabs, ragged2e}

\newcommand{\be}{\begin{equation}}
\newcommand{\ee}{\end{equation}}
\newcommand{\bea}{\begin{eqnarray}}
\newcommand{\eea}{\end{eqnarray}}
\newcommand{\bear}{\begin{eqnarray}}
\newcommand{\eear}{\end{eqnarray}}
\newcommand{\ba}{\begin{array}}
\newcommand{\ea}{\end{array}}
\newcommand{\Pb}{{\cal P}}
\newcommand{\Sp}{\Phi}

\newcommand{\prm}{{\rho}}
\renewcommand{\arraystretch}{1.3}
\begin{document}

\baselineskip=18pt \pagestyle{plain} \setcounter{page}{1}

\vspace*{-1.cm}

\noindent \makebox[11.9cm][l]{\small \hspace*{-.2cm} }{\small Fermilab-Pub-25-0021-T}  \\  [-1mm]

\begin{center}

{\bf  \normalsize 
\Large  
Composite quarks and leptons with embedded QCD 
} 

\vspace*{1.cm}

{\normalsize \bf 
Beno\^it Assi$^{1,2}$ and Bogdan A. Dobrescu$^1$} \\ [4mm]
{\small \it $^1$ Particle Theory Department, Fermilab, Batavia, Illinois 60510, USA     }  \\ [1mm]
{\small \it $^2$ Department of Physics, University of Cincinnati, Cincinnati, Ohio 45221, USA     }  \\ [4mm]

\vspace*{0.6cm}

January 17, 2025; revised October 5, 2025

\end{center}

\vspace*{0.6cm}

\begin{abstract} \normalsize
We construct a model of quark and lepton compositeness based on an $SU(15)$ gauge interaction that confines chiral preons, which are also charged under the weakly-coupled $SU(4)_{\rm PS} \times SU(2)_L\times SU(2)_R$ gauge group. The breaking of the latter, down to the Standard Model group, is achieved by scalar $SU(15)$ bound states at a scale in the $30 - 100$ TeV range. The embedding of the QCD gauge group in $SU(4)_{\rm PS} $ slows down the running of $\alpha_s$ in the UV. We estimate the effects of the strongly-coupled $SU(15)$ dynamics on the running of the $SU(4)_{\rm PS}  \times SU(2)_L\times SU(2)_R$ gauge couplings, which likely remain perturbative beyond the compositeness scale of about $10^3 - 10^4$ TeV, and even above a unification scale. A composite vectorlike lepton doublet acquires a mass in the TeV range probed at future colliders, and an extended Higgs sector arises from 6-preon bound states. 
\end{abstract}

\vspace*{-1cm}

\newpage

\renewcommand*\contentsname{\small\normalsize\large Contents}
{\small\normalsize \tableofcontents}

\bigskip \bigskip

\section{Introduction}
\label{sec:intro}

The quest for uncovering deeper building blocks of matter was at the heart of physics research throughout the 20th century, and yielded outstanding scientific breakthroughs. 
Since the beginning of this century, that quest has been wavering, probably due to the inherent difficulties of making progress towards understanding strongly-coupled quantum field theories with chiral fermions \cite{tHooft:1979rat}.
Chiral gauge theories appear to be necessary for describing the quarks and leptons as bound states, given that all Standard Model (SM) fermions have chiral charges under the electroweak gauge group.

A recently proposed theory of quark and lepton compositeness \cite{Dobrescu:2021fny} 
attempts to describe the observed properties of the quarks and leptons based on an $SU(15)_{\rm p}$ gauge interaction that confines an anomaly-free set of chiral fermions, referred to as preons (for historical reasons,  {\it e.g.},  \cite{Pati:1980rx,Bars:1981se,Geng:1986xh}). The latter include 15 fermions charged (the same way as a single generation of SM  fermions) under the SM gauge group, $SU(3)_c \times SU(2)_W \times U(1)_Y$, which does not play a role in confining the preons.
That theory has several successful features: it implies the existence of three generations of composite quarks and leptons with the SM gauge charges, it provides a dynamical origin for the SM Higgs doublet as a 6-preon bound state, and it predicts the existence of certain composite vectorlike quarks and leptons that could be searched for at the LHC and future colliders.

At the same time,  quark and lepton compositeness based on $SU(15)_{\rm p}$ is facing some challenges \cite{Assi:2022jwg}. First, dimension-8 proton decay operators are likely induced at the compositeness scale, such that the latter must satisfy $\Lambda_{\rm pre} \gtrsim 10^3$ TeV, with  uncertainties of more than an order of magnitude due to unknown nonperturbative effects.
Second, the presence of several composite vectorlike quarks implies that the
QCD coupling loses asymptotic freedom above a scale of about 10 TeV, and barely remains perturbative up to $\Lambda_{\rm pre}$. This suggests that $SU(3)_c$ should be 
embedded in a larger group, so that the gauge coupling has a smaller positive  $\beta$ function ({\it i.e.}, slower running), or even a negative one  
({\it i.e.}, asymptotic freedom is recovered at higher scales).

Moreover, due to the composite vectorlike quarks and leptons, the $U(1)_Y$ coupling loses perturbativity at a scale above $\Lambda_{\rm pre}$ but far below the GUT scale. 
Although we cannot reliably predict what occurs in the region near the scale $\Lambda_{\rm pre}$ where the $SU(15)_{\rm p}$ interaction is strongly coupled, 
a self-consistent description of the  hypercharge gauge interaction requires embedding $U(1)_Y$ in a non-Abelian group. Only in that case  the hypercharge   $\beta$ function becomes smaller or even negative, due to the negative contribution from the gauge boson self-coupling. 

For the above reasons, in this paper we study a preonic $SU(15)_{\rm p}$  theory in which the preons are also charged under the minimal non-Abelian extension of the SM gauge group: $SU(4)_{\rm PS} \times SU(2)_L \times SU(2)_R$. While the latter was proposed in \cite{Pati:1974yy} to provide a unified origin for quarks and leptons, we study it here as a weakly-coupled non-Abelian gauge group acting on  preons as well as on preonic bounds states. Embedding the QCD gauge group in $SU(4)_{\rm PS}$ has the advantage that its gauge coupling runs slower at scales above the masses of the composite vectorlike quarks. Moreover, we show that the binding due to the additional gauge bosons may push the masses of several composite vectorlike fermions towards the $\Lambda_{\rm pre}$ scale, further softening the running of the gauge couplings. 

We also show that the composite scalars formed as di-prebaryon states (akin to the deuteron in QCD) may provide the origin of $SU(4)_{\rm PS} \times SU(2)_R$ breaking down to $SU(3)_c \times U(1)_Y$ at a scale $\Lambda_{\rm PS} < \Lambda_{\rm pre}$. While robust proofs of the mechanisms outlined here are not possible in the absence of a better understanding of strongly-coupled chiral gauge theories, which so far eludes lattice field theory (for recent attempts, see \cite{Kaplan:2023pxd}), the $SU(15)_{\rm p} \times SU(4)_{\rm PS} \times SU(2)_L \times SU(2)_R$ gauge theory proposed here appears to have the necessary ingredients for a realistic theory of quark and lepton compositeness.

The paper is organized as follows. In Section~\ref{sec:model} we introduce the preonic model above the compositeness scale. In Section~\ref{sec:below} we study the effective theory below the compositeness scale.  In particular, we examine the spontaneous breaking of $SU(4)_{\rm PS} \times SU(2)_R$ due to composite dynamics, followed by an analysis of the theory at scales below $\Lambda_{\rm PS}$. In Section~\ref{sec:beta}, we compute within this framework the running of the three gauge couplings  up to the compositeness scale, and then investigate how the gauge couplings may evolve above $\Lambda_{\rm pre}$. We summarize our findings and provide an outlook in Section~\ref{sec:conclusion}.

\bigskip

\section{Preonic theory}
\label{sec:model}\setcounter{equation}{0}

The theory proposed in this paper has the same preonic gauge group as the one analyzed in \cite{Dobrescu:2021fny, Assi:2022jwg}, but crucially it has a larger weakly-coupled gauge group, which we will argue later on (Section 3) that it is spontaneously broken down to the SM gauge group by the dynamics of the preonic gauge interaction.

\subsection{Field content}

We consider a preonic theory with the $SU(15)_{\rm p} \times SU(4)_{\rm PS} \times SU(2)_L\times SU(2)_R $ gauge symmetry. The $SU(15)_{\rm p}$ group becomes strongly coupled at a scale $\Lambda_{\rm pre}$, a few orders of magnitude above the electroweak scale.
$SU(4)_{\rm PS}$ is the weakly-coupled group that embeds QCD, $SU(2)_L$ is the SM gauge group associated with the weak interaction,
and $SU(2)_R $ is the weakly-coupled gauge group that together with the $U(1)_{B-L}$ subgroup of $SU(4)_{\rm PS}$ eventually breaks down to the SM hypercharge gauge group. 

The preons are chiral fermions which transform nontrivially under $SU(15)_{\rm p}$: several of them 
($\Psi_W , \Psi_{W'}, \psi_1, \psi_2, \psi_3$) belong to the fundamental representation, and one fermion ($\Omega$) belongs to the  conjugate symmetric representation. Only the $\Psi_W$ and $\Psi_{W'}$ fields are charged under $SU(4)_{\rm PS} \times SU(2)_L\times SU(2)_R $,  as shown in Table~\ref{table:preons}. This set of six chiral fermion representations is free of gauge anomalies. In particular, the $SU(15)_{\rm p}$ gauge anomaly of $\Omega$ is cancelled by 19 fermions transforming in the fundamental representation \cite{Eichten:1982pn}; this is achieved by the  $\psi_1, \psi_2, \psi_3$ fermions together with the 16 fundamentals of $SU(15)_{\rm p}$  packaged in $\Psi_W$ and $\Psi_{W'}$. 

The only other elementary fields present at scales of the order of $\Lambda_{\rm pre}$ are two scalars ($\cal A$, $\cal A'$)  in the conjugate antisymmetric representation of $SU(15)_{\rm p}$, whose Yukawa couplings are ultimately responsible for the flavor dependence of the SM fermion masses (see Section~\ref{sec:belowPS}).  The Yukawa couplings of the antisymmetric scalars  to preons can be written as
\be
\mathcal{L}_{\rm Yuk}=\lambda_{ij} \,  {\cal A}_{ab}  \, \psi_i^a  \psi_j^b + \lambda_{ij}^\prime \,  {\cal A}'_{ab}  \, \psi_i^a  \psi_j^b    + {\rm H.c.}   ~~,
\label{eq:Apsipsi}
\ee
where $i,j = 1,2,3$ are preonic flavor indices, and $a,b = 1, ...,15$  are $SU(15)_{\rm p}$-color indices.
The Yukawa couplings $\lambda_{ij}$ form a complex symmetric matrix. Through an appropriate redefinition of the fermion fields—implementing a Takagi factorization—this matrix can be diagonalized with all entries ($\lambda_i$) real. In that field basis, $\lambda_{ij}^\prime$ generically form a complex symmetric matrix with both off-diagonal and diagonal nonzero  entries.

\begin{table}[t!]
\begin{center}
\renewcommand{\arraystretch}{1.5}
\begin{tabular}{|c|c|c|c |}\hline  
  field   &    \,  spin  \,  &  $SU(15)_{\rm p}$   &  $ SU(4)_{\rm PS} \times SU(2)_L\times SU(2)_R $    
\\ \hline \hline
    $ \Psi_W $   &  1/2  &    $\Box$    &   $(\, 4 \, , \,  2 \, , \, 1 \, )$    
 \\   \hline  
      $ \Psi_{W'} $   &  1/2  &    $\Box$    &   $(\, \bar 4 \, , \, 1 \, , \, 2 \, )$     
 \\   \hline  
   \,    $ \psi_1,  \psi_2 ,   \psi_3     $  \,    &  1/2  &   $\Box$  &  $( \, 1  \, ,  \, 1  \, ,  \, 1  \, ) $     
 \\  \hline   
      $\Omega$   &  1/2  &    $\overset{  \overset{ \rule{3.2ex}{0.07em} }{ {} }  }{  \Box\!\Box  } $   &   $( \, 1  \, ,  \, 1  \, ,  \, 1  \, ) $   
   \\  \hline  \\[-7mm] 
   $\cal A$  , $\cal A'$   &  0    &   $\overset{  \overset{ \rule{1.6ex}{0.07em} }{ {} }  }{  \parbox[c]{0.3cm}{$\Box$ \\[-2.95mm]  $\Box$ \\ [-4.5mm] }  }$  &  $( \, 1  \, ,  \, 1  \, ,  \, 1  \, ) $   
\\   \hline 
\end{tabular}
\caption{Field content of the preonic model. All fields (6 chiral fermions and two scalars) belong to nontrivial representations of the confining $SU(15)_{\rm p}$ group.
Only the $ \Psi_W $ and $ \Psi_{W'} $ preons carry charges of the weakly-coupled $SU(4)_{\rm PS} \times SU(2)_L\times SU(2)_R$ gauge group. 
 \\[-0.7cm] }
\label{table:preons}
\end{center}
\end{table}

\smallskip

\subsection{Preonic dynamics}
\label{sec:dynamics}

Given that the second Dynkin indices for the symmetric and the antisymmetric  $SU(15)$ representations are $T_2(\Omega) =17/2$ and $T_2( {\cal A} ) = 13/2$, respectively \cite{Eichten:1982pn},  we find 
the following 1-loop coefficient of the $SU(15)_{\rm p}$ $\beta$ function:
\be
b_{15} = 55 
- \frac{2}{3} \left(\frac{19}{2} +  T_2(\Omega) \rule{0mm}{3.9mm}   \right) 
- \frac{2}{3} \,  T_2( {\cal A} )  =  \frac{116}{3} 
~~.
\ee
Thus,  $b_{15} > 0$ ({\it i.e.}, negative $\beta$ function), implying asymptotic freedom for the $SU(15)_{\rm p}$ gauge interaction. Consequently, it is natural that $SU(15)_{\rm p}$ is weakly coupled near the Planck scale ($M_{\rm P}$), and that its gauge coupling grows logarithmically towards lower scales until it 
becomes nonperturbative at $\Lambda_{\rm pre}$.

The dynamical behavior of the $SU(15)_{\rm p}$ theory at the scale $\Lambda_{\rm pre}$ cannot be decided (for now) through a direct computation. The reason is that  this is a strongly-coupled chiral gauge interaction, and the only tool available for first-principle based computations is lattice field theory, which currently cannot be applied to chiral theories.
Nevertheless, there are consistency checks that indicate some possible dynamics. Specifically, there are three mutually-excluded proposals in the literature for the dynamical outcomes of an $SU(N)_{\rm p}$ gauge theory with chiral fermions in one conjugate symmetric 2-tensor and $N+4$ fundamental representations.

The first one, which we assume to be the correct one in Sections 3 and 4, is that there is confinement at the scale $\Lambda_{\rm pre}$, and that the chiral symmetry of the elementary fermions is unbroken. Even though the latter behavior is different from QCD, which spontaneously breaks the chiral symmetry of the quarks, it is a self-consistent dynamics \cite{Eichten:1985fs, Smith:2021vbf, Karasik:2022gve, Bars:1981se, Appelquist:1999vs, Appelquist:2000qg} supported by both the 't Hooft anomaly matching condition and large $N$ arguments \cite{Eichten:1985fs, Karasik:2022gve}. 

Second proposal is that there is no confinement, and instead there are chiral condensates that break completely the $SU(N)_{\rm p}$ gauge symmetry \cite{Appelquist:2000qg}. Intriguingly, the spectrum of massless fermions is similar to the case of confinement discussed above (almost consistent with the conjecture of complementarity \cite{Dimopoulos:1980hn, Geng:1986xh}). In \cite{Appelquist:2000qg} it is argued that confinement and no chiral symmetry breaking is a more likely scenario than the formation of chiral condensates, based on a conjecture regarding the thermodynamic free energy applied to chiral gauge theories \cite{Appelquist:1999vs}. In \cite{Bolognesi:2021hmg}, the opposite is argued based on discrete symmetry matching; however, that argument was disputed in  \cite{Smith:2021vbf} (for follow-ups, see \cite{Karasik:2022gve, Bolognesi:2021jzs}).

Third proposal is that there is confinement, but chiral condensates partially break the $SU(N)_{\rm p}$ gauge symmetry \cite{Csaki:2021aqv}. This is based on supersymmetric dualities perturbed by small anomaly-mediated supersymmetry breaking terms. 
As emphasized in \cite{Karasik:2022gve,Csaki:2021aqv}, for larger supersymmetry breaking it is possible that a phase transition occurs, and the result obtained in  \cite{Csaki:2021aqv} might not apply to the non-supersymmetric theory. In fact, some evidence that this is the case is provided by the large $N$ arguments: the phase identified in \cite{Csaki:2021aqv} is valid for {\it any} $N \ge 13$, while the derivation of the unbroken chiral symmetry is robust at $N \to \infty $ \cite{Eichten:1985fs, Karasik:2022gve}. In our case, $N=15$ seems large enough to satisfy the large $N$ arguments, but for now other phases cannot be conclusively ruled out. 

It is worth emphasizing that the theory studied here has additional ingredients compared to $SU(N)_{\rm p}$ theory with a symmetric tensor and $N+4$ fundamental representations: there are two antisymmetric scalars (see Table~\ref{table:preons}) with masses presumably of the order of or slightly  above $\Lambda_{\rm pre}$, and there are three other gauge groups that are weakly-coupled at $\Lambda_{\rm pre}$. We will argue in Section \ref{sec:below} that their effects may be sufficient to induce small VEVs for composite scalars, and masses for the composite quarks and leptons.

Turning now to the running of the weakly-coupled gauge couplings in the presence of preons (the runinng at scales below $\Lambda_{\rm pre}$ is analyzed in Section \ref{sec:beta}), the 1-loop coefficient of the $\beta$ function is $b_{4} = - 16/3$ for $SU(4)_{\rm PS}$, and $b_{2} = - 38/3$ for both $SU(2)_L$ and $SU(2)_R$. As $b_{4}, b_{2} < 0$, the $SU(4)_{\rm PS} \times SU(2)_L\times SU(2)_R$ gauge interactions are not asymptotically free. Furthermore, the large values of $|b_4|$ and $|b_2|$ indicate that these interactions could become strongly coupled at a UV scale below $M_{\rm P}$.  Note though that $|b_4|$ is smaller than the analogous quantity for QCD, if $SU(3)_c$ were not embedded in $SU(4)_{\rm PS}$. Therefore, the embedding pushes the UV scale to higher values. The same is true for the embedding of the hypercharge gauge group in $SU(4)_{\rm PS}\times SU(2)_R$.

There are, however, mitigating effects due to $SU(15)_{\rm p}$ dynamics near $\Lambda_{\rm pre}$ (see Section \ref{scn:premesons}) that decrease the $SU(4)_{\rm PS} \times SU(2)_L\times SU(2)_R$  gauge couplings such that they may remain weakly coupled up to $M_{\rm P}$.  
As it is difficult to estimate the size of these effects, we note that if the couplings become  strongly coupled at a scale below $M_{\rm P}$, then either there is some strongly-coupled UV fixed point, or $SU(4)_{\rm PS} \times SU(2)_L\times SU(2)_R$ must be embedded in a larger gauge group at a scale $\Lambda_{422}$ that satisfies  $\Lambda_{\rm pre}  < \Lambda_{422} < M_{\rm P}$. That larger group can be a grand unified one, such as $SO(10)$,  or just a product of groups with higher rank than $SU(4) \times SU(2)^2$. 

\smallskip

\section{Bound states of $SU(15)_{\rm p}$} 
\label{sec:below} \setcounter{equation}{0}

At the scale $\Lambda_{\rm pre}$, the $SU(15)_{\rm p}$ interactions become strong and form $SU(15)_{\rm p}$-singlet bound states. 
The chiral symmetry of the preons remains unbroken, so some fermionic bound states, referred to as  chiral prebaryons,
remain massless at this stage. The 't Hooft anomaly matching conditions \cite{tHooft:1979rat} indicate that the only chiral prebaryons are 
3-preon states 
formed of one preon belonging to the $SU(15)$ conjugate symmetric tensor representation ($\Omega$ in our model)  and two preons belonging to the fundamental representation ($\Psi_W$, $\Psi_{W'}$, or $\psi_i$, $i = 1,2,3$) \cite{Dimopoulos:1980hn, Eichten:1985fs}.  Thus, the following 3-preon states (which are 2-component fermions) remain massless:
$\Omega  \, \Psi_W \, \psi_i \, $,  \ $\Omega  \, \Psi_{W'}  \, \psi_i$ , $\Omega \, \psi_i \, \psi_j$ , $\Omega  \, \Psi_W \, \Psi_{W'}$, $\Omega  \, \Psi_W \, \Psi_W$ , $\Omega  \, \Psi_{W'} \, \Psi_{W'}$,   where $i,j = 1,2,3$ and $i \neq j$.  
Since the $\Psi_W$ and $\Psi_{W'}$ preons are charged under the weakly-coupled gauge group (see Table~\ref{table:preons}), the  chiral prebaryons belong to the $SU(4)_{\rm PS} \times SU(2)_{\rm L}  \times SU(2)_{\rm R} $ representations displayed in Table~\ref{table:PSprebaryons}.

\begin{table}[t!]
\begin{center}
\renewcommand{\arraystretch}{1.5}
 \vspace{0.2cm}
\begin{tabular}{|c|c|c|}\hline    
 \ prebaryon \ & \ preon content \  &   $SU(4)_{\rm PS} \times SU(2)_{\rm L}  \times SU(2)_{\rm R} $
\\ \hline \hline
 $\Pb_{15}$  ,  $\Pb_{1, 2}$     &      $ \Omega  \, \Psi_W   \, \Psi_{W'}  $   &  $ (15 , 2, 2)  \,\, , \,\,  (1 , 2, 2) $  
\\ \hline
 $\Pb_{10}$ ,  $\Pb_{6,3}$  &      $\Omega \, \Psi_W  \,  \Psi_W $    &     $(10,1,1)  \,\, , \,\,    (6,3,1)$ 
\\ \hline
  $\Pb_{\overline{10}}$ ,  $\Pb_{6,1}$  &       $\Omega \, \Psi_{W'}  \,  \Psi_{W'} $    &     $(\overline{10}, 1, 1)  \,\, , \,\,   (6, 1, 3)$ 
 \\ \hline 
   $\Pb_4^i$   &       $    \Omega \, \Psi_{W}  \,   \psi_i  $     &  $ 3 \times (  4 , 2, 1) $       
\\ \hline 
 $\Pb^i_{\bar{4}}$  
 &        $    \Omega \, \Psi_{W'}  \,  \psi_i  $     &  $ 3 \times (  \overline{4} , 1, 2) $       
\\ \hline 
 $\Pb_{1,1}^{ij}$  &        $    \Omega \, \psi_i  \,  \psi_j $      &  $3 \times ( 1, 1, 1) $       
\\ \hline
 \end{tabular}
\caption{Chiral prebaryons of $SU(15)_{\rm p}$, their preon content, and their representations under the weakly-coupled gauge  group.
The preon flavor index is $i=1,2,3$, so there are three $\Pb_{4}^i$ prebaryons and three $\Pb_{\bar 4}^i$  prebaryons. The anticommutation property of the preons implies that 
the $\Pb_{1,1}^{i j}$  prebaryons are antisymmetric in the $i,j$ indices, so there are also three of them.  \\ [-9mm]
}
\label{table:PSprebaryons}
\end{center}
\end{table}

We label the chiral prebaryons  as $\Pb$ with a lower index that specifies its $SU(4)_{\rm PS} $ representation, 
and when this does not uniquely determines the prebaryon, the $SU(2)_{\rm L} $ representation is also displayed after a comma, as shown in Table~\ref{table:PSprebaryons}.
For example, both $\Pb_{6,3}$ and $\Pb_{6,1}$ belong to the 6 of $SU(4)_{\rm PS} $, but the first prebaryon is a triplet under $SU(2)_{\rm L}$ while the second one is an $SU(2)_{\rm L}$ singlet. In the case of prebaryons that have one or two $\psi_i$ preons as constituents, one or two flavor upper indices ($i,j = 1,2,3$) are also included. 

At scales below the $SU(4)_{\rm PS} \times SU(2)_{\rm L}  \times SU(2)_{\rm R} $ breaking,  the  $\Pb_4^i$ and $\Pb_{\bar 4}^i$ prebaryons include 3 generations of SM quarks and leptons (see discussions in Sections \ref{sec:breaking} and \ref{sec:belowPS}). There are also 6 SM-singlet fermions, three in $\Pb_{\bar 4}^i$ and three in the $\Pb_{1,1}^{ij}$ prebaryons, which mix with the SM neutrinos. The remaining prebaryons, listed in the first three rows of Table~\ref{table:PSprebaryons},  are vectorlike under the SM gauge group, and acquire masses much larger than the electroweak scale, as discussed below.

\subsection{Scalar di-prebaryons}
\label{sec:dipre-singlet}

As the chiral symmetry of the preons is preserved, there are no light pion-like states bound by the $SU(15)_{\rm p}$ interactions. We expect, however, that bound states of two or more chiral prebaryons form due to remnant  interactions of  $SU(15)_{\rm p}$, somewhat similar to nuclear interactions in QCD. Scalars lighter than  $\Lambda_{\rm pre}$ likely arise as bound states of two chiral prebaryons \cite{Dobrescu:2021fny}, and are referred to as di-prebaryons.  Let us model the interaction between chiral prebaryons due to $SU(15)_{\rm p}$ remnant effects as 4-fermion contact terms in the Lagrangian:
\be
- \frac{g_\rho^2}{\Lambda_{\rm pre}^2}(\overline\Pb \sigma^\mu \Pb)(\overline\Pb \sigma_\mu \Pb)  ~~,
\label{eq:4Pop}
\ee
where the $\sigma^\mu$ matrices account for spin-1 premeson exchange between prebaryon Weyl spinors, $g_\rho$ is a premeson coupling 
to the prebaryons, and we have not displayed the prebaryon indices. Additional operators involving four chiral prebaryons, such as tensor-tensor couplings, are also expected to have important effects, but we will not further discuss them. 

A Fierz transformation of (\ref{eq:4Pop}) gives operators of the type $(\overline\Pb \, \overline \Pb)(\Pb \Pb)$, which are attractive and produce scalar $\Pb_\kappa \Pb_\lambda$  bound states (di-prebaryons), generically labeled $\phi_{\kappa{\text -}\lambda} $. Here $\kappa$ and $\lambda$ are prebaryon lower indices, according to the labeling in Table~\ref{table:PSprebaryons}.
Since these interactions between chiral prebaryons are nonconfining, the low-energy effective theory includes Yukawa interactions of the 
di-prebaryons to their 3-preon constituents: 
\be
- y_{\kappa{\text -}\lambda} \, \phi^\dagger_{\kappa{\text -}\lambda} \, \Pb_\kappa \Pb_\lambda ~~,
\ee
where $y_{\kappa\text-\lambda}$ are nearly-nonperturbative Yukawa couplings. 

The squared-mass $M_{\kappa\text-\lambda}^2$ of the scalar bound state $\phi_{\kappa{\text -}\lambda} $ receives  a positive contribution of the order of  $\Lambda_{\rm pre}^2$ due to the mass of the spin-1 premeson mediator, and a negative contribution of the order of  $y_{\kappa{\text -}\lambda}^2 \Lambda_{\rm pre}^2  /(2\pi^2)$ due to prebaryon loops involving the attractive Yukawa interaction. Thus, if the  effective Yukawa coupling $y_{\kappa\text-\lambda}$ has a value near (but smaller than) a critical coupling  $y_{\rm cr} \approx \pi / \sqrt{2}$, then it is possible to have $0 < M_{\kappa\text-\lambda} ^2 \ll \Lambda_{\rm pre}^2 $.
A similar phenomenon has been studied \cite{Bardeen:1993pj} in the context of top quark condensation  \cite{Bardeen:1989ds, Hill:2024qac}
and composite Higgs models \cite{Dobrescu:1997nm}, where a second-order chiral phase transition allows a large hierarchy between the scale of compositeness and the mass of the composite scalar.

Although the binding of these $\phi$ di-prebaryons is due mostly to $SU(15)_{\rm p}$ remnant interactions, there are additional contributions from 
 $SU(4)_{\rm PS} \times SU(2)_{\rm L}  \times SU(2)_{\rm R} $ gauge interactions and scalar ($\cal A$, $\cal A'$) exchange.
The additional contributions are larger when the constituent prebaryons  belong to higher  gauge representations or have large Yukawa couplings in Eq.~(\ref{eq:Apsipsi}). In those cases, the squared-mass of $\phi$ may turn negative, implying that the di-prebaryons develop VEVs.
If a di-prebaryon acquires a VEV, then its two prebaryon constituents form a Dirac fermion that acquire 
a vectorlike mass. 

The binding potential $V(r)$ between two prebaryons $\Pb_\kappa$ and $\Pb_\lambda$ induced by one gauge boson exchange can be written as
\be
V_{\kappa{\text -}\lambda}(r) \approx - \frac{1}{2r} \left(C^{\kappa{\text -}\lambda}_4 \alpha_4 + C^{\kappa{\text -}\lambda}_L \alpha_L + C^{\kappa{\text -}\lambda}_R \alpha_R \right)  ~~,
\ee
where $\alpha_4$, $\alpha_L$, $\alpha_R$ are the $SU(4)_{\rm PS} \times SU(2)_{\rm L}  \times SU(2)_{\rm R} $ coupling constants at the $\Lambda_{\rm pre}$ scale. The binding coefficients $C_4, C_L, C_R$ are given by the sum of the quadratic Casimirs of the gauge representations in which the two prebaryons transform, minus the quadratic Casimir of the  $\phi_{\kappa\lambda}$ di-prebaryon (the $\Pb_\kappa$-$\Pb_\lambda$ bound state) \cite{Raby:1979my}:
\be
C^{\kappa{\text -}\lambda}_4  = C_4(\Pb_\kappa) + C_4(\Pb_\lambda) - C_4(\phi_{\kappa{\text -}\lambda})   ~~,
\ee
and analogously for $C_L$ and $C_R$. Thus, the most deeply-bound states are gauge singlets, {\it i.e.}, $\phi_{\kappa\text-\lambda}$ with $C_4(\phi) = C_L(\phi)  = C_R(\phi)  = 0$, and are formed of prebaryons in representations with large quadratic Casimirs. In particular, the symmetric representations of $SU(4)_{\rm PS}$, namely $\Pb_{10}$ and $\Pb_{\overline{10}}$, bind with $C^{10{\text -}\overline{10}}_4 = 9$. As this binding is large, we assume that the ensuing di-prebaryon $\phi_{10{\text -}\overline{10}}$ acquires a large VEV, which induces a Dirac mass near $\Lambda_{\rm pre}$ for $\Pb_{10}$ and $\Pb_{\overline{10}}$.

The adjoint representation of $SU(4)_{\rm PS}$, $\Pb_{15}$, binds to itself and produces a gauge singlet $\phi_{15\text-15}$ with a smaller $C^{15{\text -}15}_4 = 8$.  Additional binding in this channel is due to  the $SU(2)_{\rm L}  \times SU(2)_{\rm R} $ interactions, with 
$C^{15{\text -}15}_L = C^{15{\text -}15}_R = 3/2$. However, the SM coupling constants indicate $\alpha_L \lesssim \alpha_4 /3 $ and $\alpha_R \lesssim \alpha_L /3 $ (there is some uncertainty in the running of the three gauge couplings, as discussed in Section~\ref{sec:beta}), so it follows that the $\phi_{15\text-15}$  
di-prebaryon (again a gauge singlet) has a smaller VEV than $\phi_{10{\text -}\overline{10}}$, and generates a Majorana mass for $\Pb_{15}$ that is smaller than the Dirac mass for $\Pb_{10}$ and $\Pb_{\overline{10}}$. 

The $\Pb_{6,3}$ binds to itself with $C^{6,3{\text -}6,3}_4 = 5$ and $C^{6,3{\text -}6,3}_L  =  4$, and produces a gauge-singlet $\phi_{6,3\text-6,3}$
di-prebaryon. Likewise $\Pb_{6,1}$ binds to itself with $C^{6,1\text-6,1}_4 = 5$ and $C^{6,1\text-6,1}_R  = 4$, so the $\phi_{6,1\text-6,1}$ di-prebaryon has a slightly smaller VEV than $\phi_{6,3\text-6,3}$.
The ensuing Majorana mass for $\Pb_{6,1}$  is slightly smaller than that for $\Pb_{6,3}$, which in turn is smaller than the one for $\Pb_{15}$.

Bound states of the gauge-singlet prebaryons,  $\Pb_{1,1}^{ij}$, may form depending on the strength of the Yukawa couplings of the $\cal A$ and $\cal A'$ scalars in  Eq.~(\ref{eq:Apsipsi}). For example, if $\lambda_3, \lambda'_{32} > 1$, then $\cal A$ and $\cal A'$ exchanges produce a $\Pb_{1,1}^{32}$-$\Pb_{1,1}^{32}$ scalar bound state, which may acquire a VEV. Thus, the  $\Pb_{1, 1}^{32}$  prebaryon (a gauge singlet Weyl fermion) gets a large Majorana mass, which may produce a see-saw mechanism responsible for the tiny SM neutrino masses.

\subsection{Di-prebaryons charged under $SU(4)_{\rm PS} \times SU(2)_L  \times SU(2)_R $}
\label{sec:dipre}

Besides the  di-prebaryons discussed so far, which are all gauge-singlets, it is likely that there some di-prebaryons that transform nontrivially under $SU(4)_{\rm PS} \times SU(2)_{\rm L}  \times SU(2)_{\rm R} $. As the $\Pb_{10}$, $\Pb_{\overline{10}}$, $\Pb_{15}$ (and to a lesser extent $\Pb_{6,3}$, $\Pb_{6,1}$) prebaryons are very heavy, non-singlet bound states involving two of these are expected to have a large squared mass. Bound states involving only one heavy prebaryon and a $\Pb_4^i$ or $\Pb_{\bar 4}^j$ are more likely to acquire a VEV when the binding coefficients are large enough. We will refer to these bound states as heavy-light di-prebaryons.
In particular, this is the case of the $\Pb_{15}$-$\Pb_4^i$ bound states, labeled $\phi_{15\text-4_i}$: the tensor product of the $SU(4)_{\rm PS} \times SU(2)_{\rm L}  \times SU(2)_{\rm R} $ representations in this case is $(15,2,2) \times (4,2,1) \to (4,1,2)$, so the binding coefficients are given by $C^{15{\text -}4}_4 = 4$, $C^{15{\text -}4}_L  = 3/2$, $C^{15{\text -}4}_R  = 0$. 

Given that the $\phi_{15\text-4_i}$ di-prebaryons transform in the $(4,1,2)$ representation, a VEV for one of them would lead to a phenomenologically correct breaking pattern of the gauge symmetry, as discussed in Section \ref{sec:breaking}. 
Furthermore, the symmetry breaking pattern remains correct when all three $\phi_{15\text-4_i}$  acquire VEVs because bilinear terms in the scalar potential of the type $\phi_{15\text-4_i}^\dagger \, \phi_{15\text-4_j}$ align the VEVs in the $SU(4)_{\rm PS}  \times SU(2)_{\rm R} $ space.
These bilinear terms are induced by the antisymmetric scalar $\cal A'$ running in a loop together with a $\psi_k$ preon,
and depend on the product of two different Yukawa couplings shown in (\ref{eq:Apsipsi}): 
a  $\psi_k \psi_j \cal A'$  coupling and the Hermitian conjugate of the $\psi_i \psi_k \cal A'$   coupling.

\begin{figure}[t!]
	\begin{center}
		\hspace{-1mm} \includegraphics[width=0.79\textwidth]{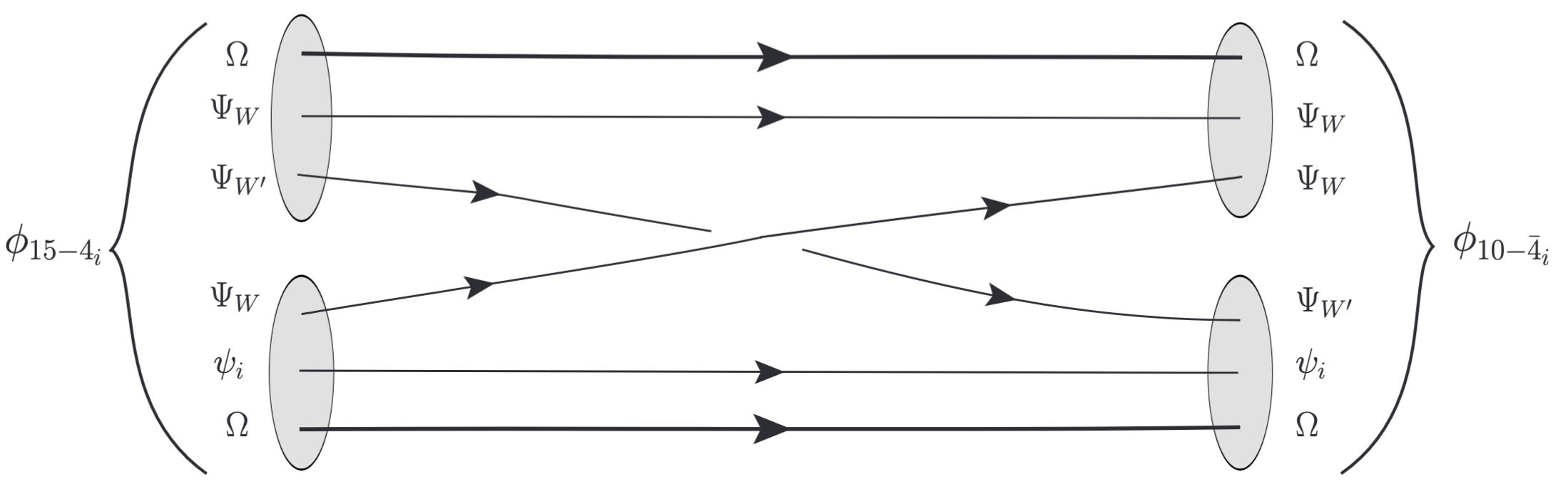} 
		\caption{Preon exchange diagram inducing the $\phi_{15\text-4_i}^\dagger \, \phi_{10\text-\bar 4_i}$ bilinear terms in the scalar potential, which are responsible for vacuum alignment along the $SU(4)_{\rm PS}  \times SU(2)_{\rm R} $ space.
		Gray bubbles represent the prebaryons (see Table \ref{table:PSprebaryons}): $\Pb_{15}$ and $\Pb_{\bar 4}^i$ form the $\phi_{15\text-4_i}$ di-prebaryons, while 
		$\Pb_{10}$ and $\Pb_{\bar 4}^i$ form the $\phi_{10\text-\bar 4_i}$ di-prebaryons.
		}
		\label{fig:bilinear}
	\end{center}
\end{figure}

In the case of the $\Pb_{15}$-$\Pb_{\bar 4}^i$ bound states, labeled $\phi_{15\text-\bar 4_i}$, the tensor product  is $(15,2,2) \times (\bar 4,1,2) \to (\bar 4,2,1) $ so 
$C^{15{\text -}\bar 4}_4 = 4$, $C^{15{\text -}\bar 4}_L= 0$, $C^{15{\text -}\bar 4}_R = 3/2$, implying weaker binding compared to $\phi_{15\text-4_i}$. 
Thus, $\phi_{15\text-4_i}$ acquire smaller VEVs than $\phi_{15\text-\bar 4_i}$, and break the electroweak symmetry because they are doublets under $SU(2)_{\rm L}$. 
The remaining heavy-light bound states are $\Pb_{10}$-$\Pb_{\bar 4}^i$ (labeled $\phi_{10\text-\bar 4_i}$) and $\Pb_{\overline{10}\,}$-$\Pb_{4}^i$  (labeled $\phi_{\overline{10}\,\text-4_i}$), which transform again as  $(4,1,2)$ and $ (\bar 4,2,1) $, respectively. Since $\Pb_{10}$ is heavier than $\Pb_{15}$, we expect that the VEVs of $\phi_{10\text-\bar 4_i}$ and $\phi_{\overline{10}\,\text-4_i}$ are smaller than those of
$\phi_{15\text-4_i}$, and could vanish. Even if all these di-prebaryons acquire VEVs, 
the symmetry breaking pattern remains correct because these VEVs are aligned with  $\langle\phi_{15\text-4_i}\rangle$ and $\langle\phi_{15\text-\bar 4_i}\rangle$ through bilinear terms in the scalar potential generated by tree-level exchange diagrams as the one shown in Figure~\ref{fig:bilinear}.

$\Pb_4^i$  and    $\Pb_{\bar 4}^j$ may also bind. Although their binding coefficient is smaller, $C^{4{\text -}\bar 4}_4 = 15/4$,  scalar exchange enhances their binding. We will assume $\lambda_3 \gg  \lambda_2, \lambda_1$, such  that ${\cal A}$ exchange in combination with premeson exchange interactions and $SU(4)_{\rm PS}$ gauge boson exchange produces a $\Pb_4^3 - \Pb_{\bar 4}^3$ bound state ($\phi_{4_3\text- \bar 4_3}$) with nonzero VEV. That  bound-state transforms as $(1,2,2)$ under $SU(4)_{\rm PS} \times SU(2)_{\rm L}  \times SU(2)_{\rm R} $, implying that under the SM gauge group it has exactly the same quantum numbers as two Higgs doublets. These provide the likely origin for SM quark and lepton masses, as discussed in Section \ref{sec:belowPS}.

\bigskip

\subsection{Spontaneous breaking of $SU(4)_{\rm PS}  \times SU(2)_{\rm R} $ }
\label{sec:breaking}

As discussed in Section \ref{sec:dipre}, there are several scalar bound states transforming in the $(4, 1, 2)$ representation of $SU(4)_{\rm PS} \times SU(2)_{\rm L}  \times SU(2)_{\rm R} $:  $\phi_{15\text-4_i}$ and $\phi_{10\text-\bar 4_i}$  for $i = 1,2,3$. We expect that some of these acquire VEVs, which are smaller than $ \Lambda_{\rm pre}$ and are aligned in the $SU(4)_{\rm PS}  \times SU(2)_{\rm R} $ space. 
There is only one linear combination of these heavy-light di-prebaryons that has a nonzero VEV, and we label it here by $\Sp$.

Since the scalar $\Sp$  transforms as $(4, 1, 2)$ under $SU(4)_{\rm PS} \times SU(2)_{\rm L}  \times SU(2)_{\rm R} $, the general renormalizable potential for  $\Sp$ is
\be
V(\Sp) = - M_{_\Sp}^2 \, {\rm Tr} \,   \Sp^\dagger \Sp +   \frac{\lambda}{2} \,    \left(  {\rm Tr} \,  \Sp^\dagger \Sp \right)^2    
+   \frac{\kappa}{2} \,    {\rm Tr} \left( \Sp^\dagger \Sp \right)^2  ~~.
\ee
We assume $M_{_\Sp}^2 > 0$ so that the potential has minima away from the origin, and for concreteness we choose $M_{_\Sp} > 0$.
The above potential is asymptotically stable for $\kappa > 0$ or  $0 \leq - 2\kappa <  \lambda/2 $.
To analyze the minima of this potential, we first use an $SU(4)_{\rm PS} \times SU(2)_{\rm R}  $ transformation to write without loss of generality the VEV of $\Sp $ as
\be
\langle \Sp \rangle = e^{i \alpha/2} \, \left( \ba{cc}    0  &  0   \\ [-1mm]  0  &  0 \\ [-1mm]  0  & s_2 \\ [-1mm]   s_1 & 0     \ea  \right)   ~~,
\ee
where $s_1, s_2 \ge 0$ have mass dimension 1, and the phase satisfies $- \pi <  \alpha \le  \pi $. 
There are two nontrivial stationarity conditions for the potential:  
\bear
&& \frac{1}{2} \,\frac{\partial V } {\partial s_1} = s_1  \left[ \left( \lambda + \kappa \right)   s_1^2 +  \lambda  \,  s_2^2  - M_{_\Sp}^2 \right]  = 0  ~~,
\nonumber \\ [-2mm]
\label{eq:first-deriv}
\\ [-2mm]
&& \frac{1}{2} \,\frac{\partial V } {\partial s_2} = s_2  \left[ \left( \lambda + \kappa \right)   s_2^2 +  \lambda  \,  s_1^2  - M_{_\Sp}^2 \right]  = 0  ~~.
\nonumber 
\eear 
The case where $s_2 = 0$ and $s_1 \neq 0$ gives
\be  
s_1 = \frac{M_{_\Sp}}{\sqrt{\lambda + \kappa}} \; \; , \; \;   \left. V(\Sp)_{} \right|_{s_2 = 0} = - \frac{M_{_\Sp}^4}{2(\lambda + \kappa)}    ~~.      
\label{eq:s20}
\ee
The case where $s_1 = 0$ and $s_2 \neq 0$  is the same up to an  $SU(4)_{\rm PS} \times SU(2)_{\rm R}  $ transformation.
Stationarity conditions (\ref{eq:first-deriv}) are also satisfied for  $s_1 , s_2 \neq 0$ provided
\be
s_1 = s_2 =  \frac{M_{_\Sp}}{\sqrt{\lambda + 2 \kappa}}  \; \; , \; \;    \left.  V(\Sp)_{} \right|_{s_1 = s_2} = - \frac{M_{_\Sp}^4}{\lambda + 2\kappa}    ~~.
\ee

The global minimum of $V(\Sp)$ is the one shown in (\ref{eq:s20}) if and only if 
\be
0 < - \lambda < \kappa  ~~.
\ee
That vacuum breaks $SU(4)_{\rm PS} \times SU(2)_{\rm R}  $ down to $SU(3)_{\rm c} \times U(1)$ 
where the generator of $U(1)$ is a linear combination of the diagonal $T_{15}= {\rm diag} \, (1,1,1,-3)/ \sqrt{24} $ generator of $SU(4)_{\rm PS}$ and the diagonal generator $T_3$ of $SU(2)_{\rm R}$.  
The gauge couplings of $SU(4)_{\rm PS}$ and $SU(3)_{\rm c}$ are the same, $g_s$.
In order to find the gauge coupling of the unbroken $U(1)$, consider the kinetic term for the scalar $\Sp$ and focus on the terms that involve the generators mentioned above:
\be
(D^\mu \Sp^\dagger) (D_\mu \Sp) \supset \left|  \left( g_s G^{15 \mu}  T_{15} + g_{_R} R^{3 \mu} T_3 \right) \langle \Sp \rangle \right|^2
= \frac{ \left( 3 g_s^2 + 4 g_{_R}^2 \right) M_{_\Sp}^2 }{ 8( \lambda + \kappa)  } \, Z^{\prime \mu} Z^\prime_\mu   ~~.
\ee
Here $G^{15 \mu}$ and  $R^{3 \mu}$ are  $SU(4)_{\rm PS} $ and $SU(2)_{\rm R}  $ gauge bosons, respectively, while $Z^{\prime \mu}$ 
is their linear combination that becomes massive,
\be
Z^{\prime \mu} =  G^{15 \mu}  \cos \theta_z -  R^{3 \mu}  \sin \theta_z  ~~.
\ee
The mixing angle $\theta_z \in (0, \pi/2)$ satisfies 
\be
\tan \theta_z  = \frac{2\, g_{_R}}{\sqrt{3} \, g_s  }   ~~.
\ee

The linear combination of $G^{15 \mu}$ and $R^{3 \mu}$ that remains massless at this stage, 
\be
B^\mu = G^{15 \mu}  \sin \theta_z  +  R^{3 \mu}  \cos \theta_z   ~~,
\ee
is the SM hypercharge gauge boson. To see that, note that the interactions of this gauge boson are determined by the couplings of the $G^{15 \mu}$  and $R^{3 \mu}$  gauge bosons to the quarks and leptons contained in the  $\Pb_4^i$ and $\Pb_{\bar 4}^i$ prebaryons (see Tables~\ref{table:PSprebaryons} and \ref{table:SMprebaryons}),
which belong to the $(  4 , 1) $ and  $(  \overline{4} , 2) $  representations of $SU(4)_{\rm PS} \times SU(2)_{\rm R}$: 
\be \hspace*{-0.2cm} 
\frac{g_s}{2\sqrt{6}} G^{15 \mu} \left( \bar q_L \gamma^\mu q_L - 3 \bar \ell_L \gamma^\mu \ell_L \right) =
g_Y B^\mu  \left( Y_{q_L} \, \bar q_L \gamma^\mu q_L +  Y_{\ell_L} \,  \bar \ell_L \gamma^\mu \ell_L \right)
+ g_z Z^{\prime \mu} J'_{L \, \mu}   ~~,
\label{eq:LH}
\ee
\bear
&&  \hspace*{-0.8cm}  \frac{g_s}{2\sqrt{6}}  G^{15 \mu} \left( \bar q_R \gamma^\mu q_R - 3 \bar \ell_R \gamma^\mu \ell_R \right) + \frac{g_{_R}}{\sqrt{2}}  R^{3 \mu}  \left( \bar u_R \gamma^\mu u_R -  \bar d_R \gamma^\mu d_R +  \bar N_R \gamma^\mu N_R -  \bar e_R \gamma^\mu e_R \right) 
\nonumber \\ [2mm]
&& \hspace*{1.8cm} 
= g_Y B^\mu  \left( Y_{u_R}  \bar u_R \gamma^\mu u_R + Y_{d_R} \bar d_R \gamma^\mu d_R + Y_{e_R}   \bar e_R \gamma^\mu e_R \right)  + g_z Z^{\prime \mu} J'_{R \, \mu}  ~~.
\label{eq:RH}
\eear
Here we are using right-handed quark and lepton field, which are the conjugates of the $\Pb_{\bar 4}^i$ prebaryons (all prebaryons in Table~\ref{table:PSprebaryons} are left-handed fermion fields).

From (\ref{eq:LH}) follows that the relation between the gauge couplings for $B^\mu$ and for gluons can be written as
\be
g_Y =  \sqrt{ \frac{3}{2} } \, g_s  \sin \theta_z ~~.
\ee
Eqs.~(\ref{eq:LH}) and (\ref{eq:RH}) are satisfied provided 
\be
Y_{q_L} = \frac{1}{6}  \;\;  ,  \;\;   Y_{\ell_L} = - \frac{1}{2}  \;\;  ,  \;\;  Y_{u_R} = \frac{2}{3}  \;\;  ,  \;\;  
Y_{d_R} = - \frac{1}{3}  \;\;  ,  \;\;  Y_{e_R} = - 1  \;\;  ,  \;\;  
\ee
which confirms that $B^\mu$ is the SM $U(1)_Y$ gauge boson.  The $J'_{L \, \mu}$ and $J'_{R \, \mu}$ represent 
the left-handed and right-handed quark and lepton currents coupled to $Z^{\prime \mu}$, with appropriate charges, 
while $g_z$ is the gauge coupling of the heavy gauge boson.

\bigskip\bigskip

\subsection{Effective theory below the  $SU(4)_{\rm PS}  \times SU(2)_{\rm R}$ breaking scale}
\label{sec:belowPS}

At scales below $\Lambda_{\rm PS}$, where  the $SU(4)_{\rm PS} \times SU(2)_{\rm L} \times SU(2)_{\rm R}$ gauge symmetry is spontaneously broken down to $SU(3)_{\rm c} \times SU(2)_{\rm L} \times U(1)_Y$, the effective theory includes the composite fermions that don't form di-prebaryons with large VEVs. These include the $\Pb_{4}^i$,  $\Pb_{\bar 4}^i$, $\Pb_{1, 2}$, and  $\Pb_{1, 1}^{i1}$ prebaryons, described in Table~\ref{table:PSprebaryons}. Their SM gauge charges are shown in Table~\ref{table:SMprebaryons}.

\begin{table}[t!]
\begin{center}
\renewcommand{\arraystretch}{1.5}
\begin{tabular}{|c|c|c|c|}\hline    
 \ prebaryons \ &  preon content &   $SU(3)_{\rm c} \times SU(2)_{\rm L}  \times U(1)_Y$ & origin
\\ \hline \hline
$ q_{i_L} \, , \; \ell_{i_L}$ &    $    \Omega \,  \Psi_{W} \, \psi_i $   &  $ ( 3, 2, +1/6)   \, , \,   ( 1, 2, -1/2 )  $     &   $\Pb_{4}^i$ 
\\ \hline 
 \hspace*{-2mm} $ u^c_i \, , \, d^c_i \, , \,  e^c_i \, , \,  N^c_i $   \hspace*{-2mm} &   $    \Omega \, \Psi_{W'} \, \psi_i $        
&   \hspace*{-2.5mm}  $ (  \overline{3} , 1, - 2/3) , (  \overline{3} , 1, + 1/3) , (  1 , 1, + 1) , ( 1 , 1, 0)  $      \hspace*{-2.5mm}                    &  $\Pb_{\bar 4}^i$ 
\\ \hline 
   $N_{31} \, , \,  N_{21} $ &   \hspace*{-2mm} $\Omega \,  \psi_3  \psi_1 $ ,  $    \Omega \,  \psi_2  \psi_1 $   \hspace*{-3mm}  
    &  $( 1, 1, 0) $                   &   \hspace*{-2mm}  $\Pb_{1, 1}^{31}$ , $\Pb_{1, 1}^{21}$  \hspace*{-2mm} 
\\ \hline
${\cal L}^c$,  ${\cal L}_{_L}$  &    $ \Omega \, \Psi_W  \Psi_{W'}  $   &  $ (1 , 2,  +1/2) + (1 , 2,  -1/2) $    &  $\Pb_{1, 2}$ 
\\ \hline    
 \end{tabular}
 \vspace{0.1cm}
\caption{Prebaryons of $SU(15)_{\rm p}$ lighter than $\Lambda_{\rm PS} $ and their representations under the SM  gauge  group.
The SM quarks and leptons arise as $\Omega  \Psi_{W}  \psi_i $ and $\Omega  \Psi_{W'}  \psi_i$ prebaryons, with the generation index $i =1,2,3$ identified as the preon flavor index. Fermions beyond the SM (below $\Lambda_{\rm PS} $) include five SM singlets and one vectorlike lepton doublet. Last column includes the original prebaryons, before $SU(4)_{\rm PS}  \times SU(2)_{\rm R} $ breaking  (see Table \ref{table:PSprebaryons}).
}
\label{table:SMprebaryons}
\end{center}
\end{table}

\bigskip

\subsubsection{SM fermions and composite right-handed neutrinos}

The prebaryons with preon content $\Omega \, \Psi_W \, \psi_i$ represent the left-handed doublets of the SM:
the $q_{i L}$ quarks and $\ell_{i L}$ leptons.  The index $i = 1,2,3$ labels the three generations, arising from the flavor index of the preons $\psi_i$. 
The prebaryons with preon content  $\Omega \, \Psi_{W'} \, \psi_i$ represent the charge-conjugates of the right-handed SM fermions: 
up-type quarks $u^c_i$, down-type quarks $d^c_i$, and charged leptons $e^c_i$. In addition, the three conjugates of right-handed neutrinos, $N^c_i$, also arise  as $\Omega \, \Psi_{W'} \, \psi_i$ bound states; they are gauge singlets under the SM gauge group, but couple to the heavy gauge bosons associated with the  breaking  of $SU(4)_{\rm PS}  \times SU(2)_{\rm R}$.

The $\Pb_{1, 1}^{ij}$ prebaryons, whose preon content is $\Omega \, \psi_i \, \psi_j$ (with $i \neq j$),  are neutral not only under the SM gauge interactions but also under all $SU(4)_{\rm PS} \times SU(2)_{\rm L} \times SU(2)_{\rm R}$ gauge symmetries.
Following the discussion at the end of Section~\ref{sec:dipre}, we assume that $\Pb_{1, 1}^{32}$ has a Majorana mass above $\Lambda_{\rm PS}$, due to $\cal A$-exchanges; 
this can be responsible for a high-mass seesaw mechanism involving the SM neutrinos. The other two $\Pb_{1, 1}^{ij}$ prebaryons, which for brevity we label here $N_{12}$ and $N_{13}$, have a Dirac mass below $\Lambda_{\rm PS}$; nevertheless, they likely have some mixings with the $N^c_i$ fermions as well as with the SM neutrinos, and may mediate additional seesaw contributions to the neutrino masses. 

Hence, the three $N^c_i$ fermions play the role of sterile neutrinos. We will not explore here the phenomenological implications of this rich composite neutrino sector. 
It is also possible that one or two of the $N^c_i$ fermions are stable, in which case they would constitute a dark matter component. Other dark matter candidates could in principle be provided by physics above $\Lambda_{\rm pre}$.

\smallskip

\subsubsection{Composite vectorlike leptons}

\begin{figure}[t!]
	\begin{center}
		\hspace{0mm} \includegraphics[width=0.66\textwidth]{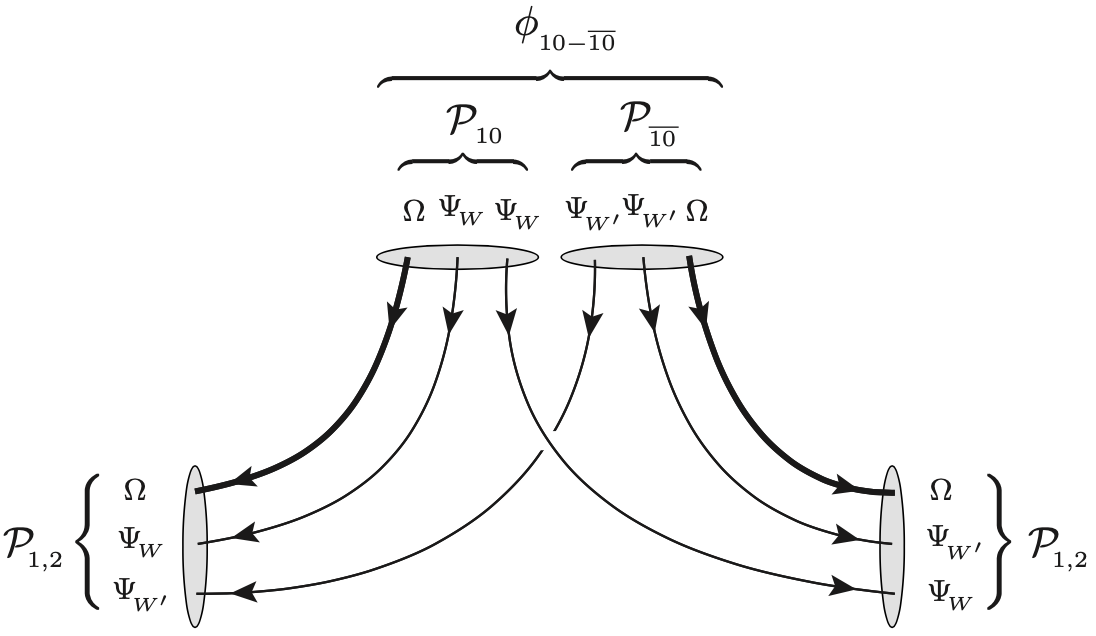} 
		\vspace*{-0.5mm}
		\caption{Preon exchange diagram inducing the effective Yukawa interaction of the $\phi_{10\text-\overline{10}}$ scalar  (a gauge-singlet bound state of the $\Pb_{10}$ and 
		 $\Pb_{\overline{10} }$  prebaryons)	 to a pair of  $\Pb_{1,2}$ prebaryons.  
		}
		\vspace*{-0.9mm}
		\label{fig:122exchange}
	\end{center}
	\begin{center}
		\hspace{0mm} \includegraphics[width=0.66\textwidth]{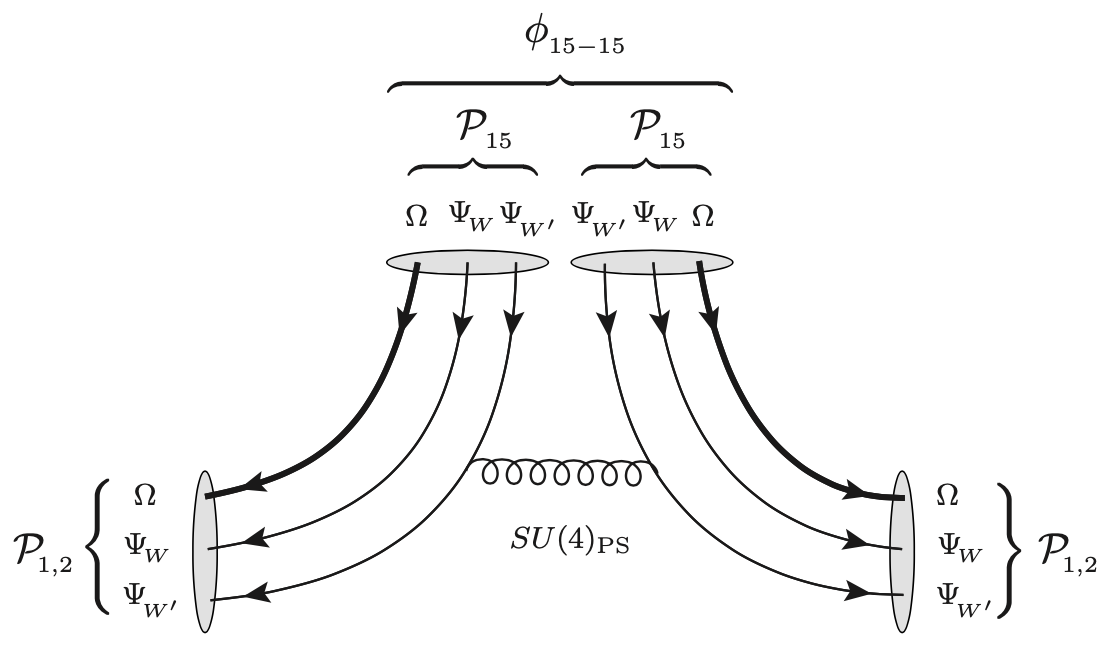} 
		\vspace*{-0.9mm}
		\caption{Effective Yukawa interaction of the  $\phi_{15\text-15}$ di-prebaryon (a gauge-singlet bound state of two $\Pb_{15}$ prebaryons)		to a pair of  $\Pb_{1,2}$ prebaryons, induced by  an $SU(4)_{\rm PS}$ gauge boson exchange. \\  [-1.cm]	}
		\label{fig:122gauge}
	\end{center}
\end{figure}

The lepton doublets ${\cal L}_L$ and ${\cal L}^c$ arise from the prebaryon $\Pb_{1,2} = \Omega\, \Psi_W\, \Psi_{W'}$. These doublets have effective Yukawa couplings to the di-prebaryons  $\phi_{10{\text -}\overline{10}}$  and $\phi_{15{\text -}15}$, which are generated by the diagrams shown in  Figures~\ref{fig:122exchange} and \ref{fig:122gauge}.
The VEVs of the di-prebaryons thus lead to to a Dirac mass $m_{\cal L}$ for ${\cal L}_L$ and ${\cal L}^c$, which form a vectorlike lepton. 
As argued in \cite{Assi:2022jwg}, the two contributions to $m_{\cal L}$ are accidentally of the same order of magnitude. The first contribution, the preon-interchange  diagram in  Figure~\ref{fig:122exchange}, is suppressed by  $1/N$ where $N = 15$ for the gauge group responsible for preon confinement.
The second contribution,  the $SU(4)_{\rm PS}$-exchange diagram in Figure~\ref{fig:122gauge}, is suppressed by $\alpha_4/\pi$. As a result, $m_{\cal L}$ may be at the TeV scale, and thus the existence of the vectorlike lepton doublet may be probed at the LHC and future colliders. 

If the vectorlike lepton doublet ${\cal L} = ( {\cal L}^0 , \cal L^- )$ mixes with the SM leptons, then the standard decay modes are  ${\cal L}^\pm \to \nu W^\pm$, $\tau^\pm Z$, $\tau^\pm h^0$  ($\nu$ generically stands here for $\nu_\tau$ or $\overline\nu_\tau$) and ${\cal L}^0 \to \tau^- W^+$, $\nu Z$, $\nu \, h^0$; note that the antiparticle of ${\cal L}^0$ then has the decay modes $\bar{\cal L}^0 \to \tau^+ W^-$, $\nu Z$, $\nu \, h^0$. 
In that case, the current LHC limit set by the  CMS  \cite{CMS:2022nty} and  ATLAS
\cite{ATLAS:2023sbu} collaborations is $m_{\cal L} > 1.0$ TeV. The High-Luminosity LHC will be sensitive to heavier vectorlike leptons.
At a $\mu^+\mu^-$ collider with center-of mass energy $\sqrt{s} > 2 m_{\cal L} $, Drell-Yan production  
of a pair of vectorlike leptons would allow precision measurements of their properties.  

In this theory of quark and lepton compositeness, 
however, the vectorlike lepton doublet may also have more exotic decay modes, especially if the pseudo-Nambu-Goldstones arising from neutral di-prebaryons have masses below  $m_{\cal L}$. If $A_{\cal L}$ is one of those pseudoscalars, then  ${\cal L}^\pm \to \tau^\pm  A_{\cal L}$ and ${\cal L}^0 \to \nu  A_{\cal L}$ may be the dominant decay modes. The production of vectorlike leptons at the LHC with the largest cross section is via an off-shell $W$ boson,  
so that the main processes are $pp \to W^{*+} \! \to {\cal L}^+ {\cal L}^0 \to (\tau^\pm  A_{\cal L})(\nu_\tau  A_{\cal L})$  and $pp \to W^{*-} \! \to {\cal L}^- \bar{\cal L}^0 \to (\tau^-  A_{\cal L})(\overline\nu_\tau  A_{\cal L})$.
The process with a $W^{*+}$ has the largest cross section (about 1 fb for $m_{\cal L} = 1$ TeV at $\sqrt{s} = 14$ TeV \cite{Bhattiprolu:2019vdu}), because the parton distribution is larger for the up quark than for the down quark.

\begin{figure}[t!]
	\begin{center}
		\hspace{0mm} \includegraphics[width=0.54\textwidth]{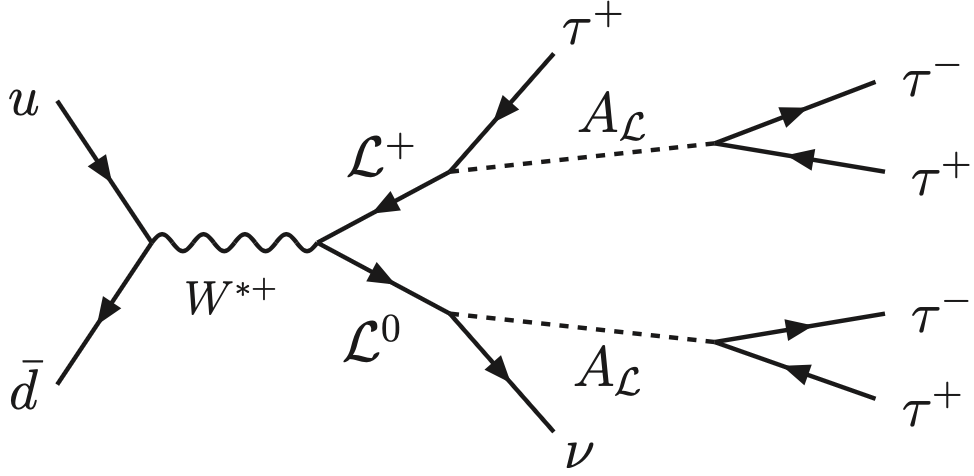} 
		\vspace*{-0.5mm}
		\caption{Main process for pair production of composite vectorlike leptons at the LHC, yielding a $5\tau \!+\! E_T \hspace*{-1.07em}\slash $ \ \ final state, provided that the mass of the composite pseudoscalar $A_{\cal L}$ satisfies $2m_\tau < M_{A} < m_{\cal L}$. 
		For $M_{A}  < 2m_\tau$, $A_{\cal L}$ predominantly decays into $\gamma\gamma$ or $\mu^+\mu^-$, so similar diagrams lead to $4\gamma + \tau \!+\!  E_T \hspace*{-1.07em}\slash $ \ \ or $4\mu + \tau \!+\!  E_T \hspace*{-1.07em}\slash $ \ \  final states.
		}
		\vspace*{-0.9mm}
		\label{fig:5tau}
	\end{center}
\end{figure}

The subsequent decays of $A_{\cal L}$ depend on its mass, and on possible couplings to the $b$ or $t$ quarks. For a $A_{\cal L}$ mass in the $3.5 - 10$ GeV,  the dominant decay is $A_{\cal L} \to \tau^+\tau^-$, which leads to the $5\tau + \nu$ final state showed in Figure~\ref{fig:5tau}. Searches in this final state have not yet been performed, and should be pursued by the  ATLAS and CMS collaborations, as well as at future collider experiments.
For $m_{\cal L} < 3.5$ GeV, the one-loop decay $A_{\cal L} \to \gamma\gamma$ competes with the $A_{\cal L} \to \mu^+\mu^-$ decay, which is suppressed by the VEV  of the neutral di-prebaryon; thus, the relevant final states are of the type $(\tau^\pm  A_{\cal L})(\nu  A_{\cal L}) \to \tau^\pm \nu + 4\gamma$ or $\tau^\pm \nu + 4\mu$, with the possibility of displaced vertices due to the long lifetime of $A_{\cal L}$.

It is interesting that the above phenomenological signatures are different from those predicted in the theory of quark and lepton compositeness \cite{Dobrescu:2021fny, Assi:2022jwg} where the weakly-coupled gauge gauge group at the $ \Lambda_{\rm pre}$ scale is the SM gauge group, instead of
$SU(4)_{\rm PS} \times SU(2)_{\rm L} \times SU(2)_{\rm R}$ as  analyzed in this paper. The lightest composite vectorlike lepton in the theory of Refs.~\cite{Dobrescu:2021fny, Assi:2022jwg} is a weak-singlet, which is produced at colliders via an off-shell $Z$ and $\gamma$, and leads to an even number of $\tau$'s in the final state \cite{Bernreuther:2023uxh}. Furthermore, in that case, there are some vectorlike quarks of charge $-1/3$ with mass potentially within the reach of the LHC.

\subsubsection{Composite Higgs sector}

As discussed in Section~\ref{sec:dipre}, the three 
$\phi_{15\text-\bar 4_i}$ di-prebaryons transform as $(\bar 4,2,1) $ and acquire VEVs smaller than the $SU(4)_{\rm PS}  \times SU(2)_{\rm R}$ breaking VEVs of 
$\phi_{15\text-4_i}$.  Given that all $SU(4)_{\rm PS}$ breaking VEVs are aligned, below $\Lambda_{\rm PS}$ each of the $\phi_{15\text-\bar 4_i}$ consists of a scalar $\tilde{q_i}^{\! c}$ transforming as  $(\overline 3, 2, -1/6)$  under the SM gauge group plus a scalar $H_i$ transforming as $(1,2, +1/2)$. Consequently there are three Higgs doublets, $H_i$ for $i = 1,2,3$, which have approximately equal VEVs (with small corrections due to ${\cal A}$ and ${\cal A}'$  loops).

Another source of electroweak symmetry breaking is the $\phi_{4_3\text-\bar 4_3}$ di-prebaryon discussed at the end of Section~\ref{sec:dipre}, which includes two Higgs doublets, $H_u$ and $H_d$. These generate the SM fermions masses through a mechanism similar to the one presented in \cite{Dobrescu:2021fny}.  The $\Pb_{4}^3$ and $\Pb_{\bar 4}^3$ prebaryons, which are identified with third-generation SM fermions as displayed in Table~\ref{table:SMprebaryons},  form the two composite Higgs doublets. Consequently, it is expected that the Yukawa couplings of third-generation SM fermions are larger than one. If the $\phi_{4_3\text-\bar 4_3}$ VEV is predominantly tilted in the direction of $H_u$, then the top quark gets a mass near the electroweak scale. Note that the Yukawa coupling larger than one is compensated by the $H_u$  VEV that is below $v \approx 246$ GeV (because the $H_i$ VEVs also contribute to electroweak symmetry breaking).
The $b$ quark and $\tau$ lepton, which are also part of the $\Pb_{4}^3$ and $\Pb_{\bar 4}^3$ prebaryons, have  Yukawa couplings to $H_d$ that are comparable to the top coupling to $H_u$, so this is a type-II two-Higgs-doublet model in the large $\tan\beta$ regime, supplanted by the three fermiophobic Higgs doublets $H_i$. 

\begin{figure}[t!]
	\begin{center}
		\hspace{0mm} \includegraphics[width=0.65\textwidth]{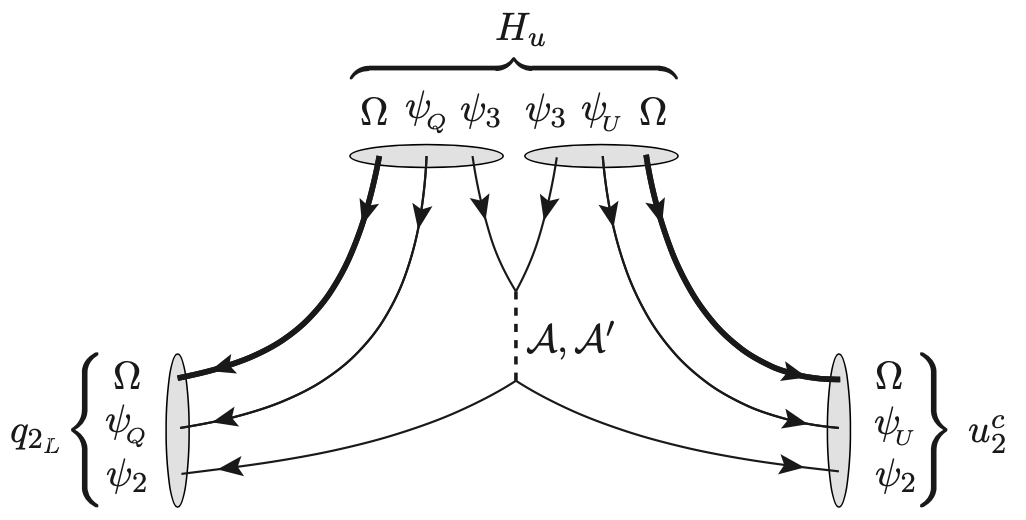} 
		\vspace*{-0.5mm}
		\caption{Effective Yukawa coupling induced by preon interactions: the $H_u$ Higgs doublet (a 6-preon bound state) couples to the second-generation SM quark doublet ($q_{2_L}$) and up-type quark singlet ($u^c_2$), producing the charm quark mass. The antisymmetric scalars ${\cal A}$ and  ${\cal A}'$ couple to different preon flavors as in Eq.~(\ref{eq:Apsipsi}). The preon $\psi_{\! _Q}$ is the color-triplet  component of $\Psi_W$, while the preon $\psi_{\! _U}$ is the color-antitriplet and upper-$SU(2)_R$ component of $\Psi_{W'}$.
		}
		\vspace*{-0.9mm}
		\label{fig:charmYukawa}
	\end{center}
\end{figure}

The SM fermions of the second and first generations have effective Yukawa couplings to $H_u$ and $H_d$ which are mediated by $\cal A$ and $\cal A'$ scalar exchange (the relevant diagram for the particular case of the charm quark is shown in Figure~\ref{fig:charmYukawa}). As a result, their masses are proportional to a sum of  products of the Yukawa couplings of $\cal A$ and $\cal A'$ shown in Eq.~(\ref{eq:Apsipsi}). 
For example, the up-type quark mass matrix ${\cal M}_u$ has the 22 element proportional to $\lambda_{33}  \lambda_{22}  +  (\lambda'_{33})^\dagger  \lambda'_{22}$. Its off-diagonal elements get contributions only from $\cal A'$ exchange in the basis where $\cal A$ has only diagonal couplings; {\it e.g.}, the 23 element of ${\cal M}_u$ is proportional to $ (\lambda'_{33})^\dagger   \lambda_{23}'$. 

Thus,  the SM quarks and leptons acquire masses, with the third generation of fermions naturally heavier than the other two, which are suppressed by a scalar exchange. 
The same would be true even if only one scalar in the antisymmetric representation of $SU(15)_{\rm p}$ existed.
However, in the absence of $\cal A'$ (as was the case in \cite{Dobrescu:2021fny}), there would not be off-diagonal Cabibbo–Kobayashi–Maskawa matrix elements unless there is an additional source of preon flavor violation (such as some higher-dimensional operators). This justifies the introduction of  $\cal A'$ in Section~\ref{sec:model}.

Given that the SM quarks and leptons obtain masses from an effective two-Higgs-doublet model of type-II, there are no tree-level flavor-changing neutral processes. There may be, however, flavor-violating higher-dimensional operators induced at the compositeness scale. We expect that their effects are adequately suppressed (given that proton stability implies $ \Lambda_{\rm pre} \gtrsim \times10^3$ TeV), but dedicated studies should analyze whether flavor processes induced by effects at the compositeness scale may eventually be observed.

\medskip \bigskip
\section{Running of gauge couplings}
\label{sec:beta}\setcounter{equation}{0}

We now explore the behavior of the gauge couplings at energy scales above the electroweak scale and up to the compositeness scale \(\Lambda_{\mathrm{pre}}\), using renormalization group equations (RGEs) \cite{Georgi:1974yf, tHooft:1973mfk, Collins:1973yy}. The evolution of a coupling $g_i$ takes the form
\begin{equation}
   \frac{dg_i}{d\ln{\mu}}=\beta_i\left(\lbrace g_j\rbrace\right) ~~,
    \label{eqn:betaDE}
\end{equation}
where $\mu$ is the renormalisation scale, and the beta function $\beta_i$ of coupling $g_i$  depends on the set of couplings denoted by $\lbrace g_j\rbrace$, which includes $g_i$. 
The beta functions are calculated perturbatively from the relationship between the bare and renormalized couplings. For a generic field  theory with gauge couplings $g_i$ and Yukawa couplings $Y^i_{jk}$ (from terms of the form $Y^i_{jk}\phi_i\Bar{\psi}_j\psi_k$ in the Lagrangian), the beta functions can be written as $\beta_i= - b_i \, g_i^3/(4\pi)^2 $ with coefficients given by
\begin{align}
   b_i= b_i^{(1)}+\frac{1}{4\pi}\left(b_i^{(2)}\alpha_i+\sum_{j\neq i}b_{ij}^{(2)}\alpha_j+\sum_j\mathcal{Y}_j\right) ~.
    \label{eqn:2loop}
\end{align}
Here $\alpha_i=g_i^2/(4\pi)$, and the one- and two-loop contributions are given by
\begin{align}
   b_i^{(1)}={}&\frac{11}{3}C(G_i)-\frac{2}{3} T_{i}(F)-\frac{1}{6}T_{i}(S) ~,
\nonumber   \\[2mm]
   b_{i}^{(2)}={}&\frac{34}{3}C(G_i)^2-C(G_i)\left(\frac{10}{3}T_i(F) + \frac{1}{3}T_i(S)\right) ~,
 \nonumber \\[-2mm]
  \label{eqn:2loop}     \\[-2mm]
   b_{ij}^{(2)}={}&-2\bigg(C_{i}(F)T_{j}(F) + C_{i}(S)T_{j}(S)\bigg) ~, 
   \nonumber  \\[2mm]
   \mathcal{Y}_{i}(F)={}&\frac{1}{d(G_i)}\mathrm{Tr} [C_i(F)Y^i(Y^i)^{\dagger}] ~.     \nonumber 
\end{align}
The quadratic Casimir invariants appearing here, $C$, correspond to those of Weyl fermions ($F$), real scalars ($S$) and gauge bosons ($G$). The second Dynkin index, $T$, refers to Weyl fermions and real scalars, and the dimension of the Lie algebra of $G$ is given by $d(G)$.

The RGEs will be for the SM gauge couplings up to the scale $\Lambda_{\rm PS}$  where there is a transition to the 
$SU(4)_{\rm PS}\times SU(2)_L\times SU(2)_R$ couplings.  The SM coupling constants associated with the $SU(3)_c\times SU(2)_W\times U(1)_Y$ gauge interactions are $\alpha_s$,  
\begin{equation}
    \alpha_2=\frac{\alpha}{\mathrm{sin}^2{\theta_W}} \, , \; \quad \alpha_Y=\frac{\alpha}{\mathrm{cos}^2{\theta_W}}   ~,
\end{equation}
where $\alpha$ is the electromagnetic coupling and $\theta_W$ is the weak mixing angle. 
 The transition between the two sets of coupling constants at the $\Lambda_{\rm PS}$  scale satisfies the following tree-level matching conditions~\cite{Mohapatra:1974gc}:
\begin{equation}
    \begin{split}
        \alpha_Y^{-1}(\Lambda_{\rm PS})={}&\alpha_{R}^{-1}(\Lambda_{\rm PS})+\frac{2}{3}\alpha_{4}^{-1}(\Lambda_{\rm PS}) ~ ,\\
        \alpha_{2}(\Lambda_{\rm PS})={}&\alpha_{L}(\Lambda_{\rm PS}) ~,\\
        \alpha_{s}(\Lambda_{\rm PS})={}&\alpha_{4}(\Lambda_{\rm PS}) ~.
    \end{split}
    \label{eqn:SMPS}
\end{equation}
One can also evolve the Yukawa couplings as they begin to play a role at the two-loop level in the running of the gauge couplings, as can be seen in Eq.~\eqref{eqn:2loop}. 

In our analysis of gauge-Yukawa theories, we have utilized two-loop order beta functions from Refs.~\cite{Machacek:1983tz,Jack:1984vj,Pickering:2001aq,Bednyakov:2021qxa},
evaluated using the packages \texttt{SARAH 4}~\cite{Staub:2013tta} and \texttt{RGbeta}~\cite{Thomsen:2021ncy}. We set initial conditions for the SM gauge couplings at the top quark mass  using \texttt{mr:C++}~\cite{Kniehl:2016enc}, as shown in Ref.~\cite{Huang:2020hdv}. 
At scales above \( m_t \), the running required recalculating beta functions at each mass threshold described in Section~\ref{sec:below}.

\subsection{Beta functions between mass thresholds} 
\label{sec:betaf}

The lightest new particles charged under the SM gauge group are the vectorlike lepton doublet (of mass $m_{\mathcal L}$), 
and the scalars associated with the second Higgs doublet (a linear combination of the states in $H_u$ and $H_d$), and the three $\phi_{15\text-\bar 4_i}$ scalars 
(which include three $\tilde q_i^c$ and three $H_i$), as discussed in Section~\ref{sec:belowPS}. We assume for simplicity that all these new particles are approximately degenerate in mass, so that the gauge coupling running up to the $m_{\mathcal L}$ scale is governed just by the SM. 
Above $m_{\mathcal L}$, the coefficients of the beta functions for the three SM gauge couplings are 
\begin{align}
    b_Y={}&-\frac{25}{3}-\frac{1 }{4\pi}\left(\frac{52 }{3} \alpha_s + \frac{27}{2} \alpha _2+\frac{245 }{18} \alpha_Y\right) ,  \nonumber \\[2mm]
    b_2={}&\frac{1}{3}-\frac{1 }{\pi} \left( 9 \alpha _s+\frac{253}{24} \alpha _2+ \frac{9}{8}\alpha_Y \right) , \\[2mm]
    b_s={}&6 +\frac{1}{4\pi}\left(4 \alpha _s-\frac{27 }{2} \alpha_2 -\frac{13}{6} \alpha_Y \right) .   \nonumber
\end{align}
Here we have included the one-loop contributions, as well as the two-loop gauge contributions. At two loops there are also contributions from the Yukawa couplings, especially those of the third generation fermions to $H_u$ and $H_d$, which  for simplicity we do not include here. 

The next threshold occurs around \(\Lambda_{\rm PS}\), where the SM gauge couplings transition to \(SU(4)_{\rm PS} \times SU(2)_L \times SU(2)_R\) couplings according to Eq.~\eqref{eqn:SMPS}. 
The following left-handed fermion representations are present here: $3\times (4,2,1)$,  $3\times (\overline 4,1,2)$, $(1,2,2)$, corresponding to the prebaryons 
$\Pb^i_4$, $\Pb^i_{\bar{4}}$,  $\Pb_{1, 2}$, respectively (see Table~\ref{table:PSprebaryons}).
The scalar representations discussed in Section~\ref{sec:dipre} are exactly the conjugates of the above fermion ones: $3\times (\overline 4,2,1)$,  $3\times (4,1,2)$, $(1,2,2)$, due to the di-prebaryons  $\phi_{15\text-\bar 4_i}$,  $\phi_{15\text-4_i}$, $\phi_{4_3\text- \bar 4_3}$, respectively. 
 Interestingly, although the underlying theory is not supersymmetric, the pair of bound states $\{ \Pb^i_4 , \phi_{15\text-\bar 4_i}^\dagger \}$ form a chiral supermultiplet for each $i = 1,2,3$, and the same is true for  $\{ \Pb^i_{\bar{4}} , \phi_{15\text-4_i}^\dagger \}$ and $\{ \Pb_{1, 2} ,  \phi_{4_3\text- \bar 4_3} \}$.
The pure gauge $\beta$ function coefficients at one and two loops, for scales between \(\Lambda_{\rm PS}\) and $m_{\Pb_6}$ (the mass of the antisymmetric prebaryons) 
are given by:
\begin{align}
    b_R={}&\frac{1}{3}-\frac{1 }{8\pi} \left(135 \alpha_4 + \frac{253}{3} \alpha_L + 9 \alpha_R  \right) , \nonumber \\[2mm]
    b_L={}&\frac{1}{3}-\frac{1 }{8\pi} \left(135 \alpha_4 + 9 \alpha_L +\frac{253}{3} \alpha_R \right) ,    \label{eq:abovePSb}  \\[2mm]
    b_4={}&\frac{26}{3}+\frac{1}{8\pi}\left(\frac{107}{3} \alpha_4 - 27 \alpha _L - 27 \alpha _R \right) . \nonumber
 \end{align}

At scales above $m_{\Pb_6}$ we include intermediate-mass states transforming under the representations \((6, 1, 3)\) and \((6, 3, 1)\), corresponding to the \(\Pb_{6,1}\) and \(\Pb_{6,3}\) prebaryons, respectively. The pure gauge coupling beta function coefficients become
\begin{align}
    b_R={}&-\frac{23}{3}-\frac{1 }{8\pi} \left(255\alpha_4 + 9 \alpha_L + \frac{1021}{3} \alpha _R \right) ,  \nonumber \\[2mm]
    b_L={}&-\frac{23}{3}-\frac{1}{8\pi}\left(255 \alpha_4 + \frac{1021}{3} \alpha_L + 9\alpha_R \right) ,   \label{eqn:betaL2} \\[2mm]
    b_4={}&\frac{14}{3}-\frac{1}{8\pi}\left(\frac{553}{3} \alpha_4+ 51 \alpha_L + 51 \alpha_R \right) .  \nonumber
\end{align}
At a scale of order $m_{\Pb_6}$ two di-prebaryons also become relevant: \(\phi_{6,1\text{-}6,1}\) and \(\phi_{6,3\text{-}6,3}\). These are gauge-singlet scalars with large Yukawa couplings to the \(\Pb_{6,1}\) and \(\Pb_{6,3}\) prebaryons. The latter belong to a higher representation of $SU(4)_{\rm PS}$ (the antisymmetric one), so their Yukawa couplings have larger two-loop effects on the gauge running compared to those  of the top and bottom quarks.  
We incorporate the effects of \(\phi_{6,1\text{-}6,1}\) and \(\phi_{6,3\text{-}6,3}\) in the numerical study presented in Section~\ref{sec:comb} 
but their effects on the gauge running will turn out to be only barely visible.

Next, we encounter the threshold corresponding to the heaviest prebaryons, which transform under the representations \((15, 2, 2)\), \((10, 1, 1)\), and \((\overline{10}, 1, 1)\), as shown in Table~\ref{table:PSprebaryons}.  
 These  prebaryons significantly impact the beta functions due to their large $SU(4)_{\rm PS}$ representations. The pure gauge beta functions at scales above $m_{\Pb_{10}}$ have the following coefficients:
\begin{align}
    b_R={}&-\frac{46}{3}-\frac{1 }{4\pi} \left(\frac{405}{2} \alpha_4 +24 \alpha_L+\frac{787 }{3} \alpha_R \right) ,  \nonumber \\[2mm]
    b_L={}&-\frac{46}{3}-\frac{1}{4\pi}\left(\frac{405}{2} \alpha_4 +\frac{787 }{3} \alpha_L +24 \alpha_R\right) ,  \label{eqn:betaL3} \\[2mm]
    b_4={}&-8-\frac{1}{8\pi}\left( 1013 \alpha_4 + 81 \alpha_L + 81 \alpha_R \right) .  \nonumber
\end{align}
These prebaryons also form two gauge-singlet scalars, \(\phi_{15\text{-}15}\) and \(\phi_{10\text{-}\overline{10}}\).  
For the purpose of calculating the running of the gauge couplings, we assume that the masses of these prebaryons and their associated scalars are degenerate at the scale $m_{\Pb_{10}}$. 
Their Yukawa couplings again contribute to the running of the gauge couplings at the two-loop order, but the effects are larger due to the higher $SU(4)_{\rm PS}$ representations. These are numerically included in Section~\ref{sec:comb}, with the starting value for the Yukawa couplings at the scale $m_{\Pb_{10}}$ being close to the nonperturbative regime, $y_{10\text-\bar 10} \approx y_{15\text-15} \approx 2$.  
 
At this stage, the presence of many composite fields has lead to the loss of asymptotic freedom for each of the $SU(4)_{\rm PS} \times SU(2)_L \times SU(2)_R$ gauge interactions.
Nevertheless, we will see
next that these interactions remain perturbative at least up to the compositeness scale.

\medskip

\subsection{Gauge coupling evolution} 
\label{sec:comb}

We perform the running with $\overline{\mathrm{MS}}$ renormalized beta functions. At the mass of the top quark, $m_t \approx 173$ GeV \cite{ParticleDataGroup:2024cfk},  the SM gauge coupling constants are given by \cite{Huang:2020hdv}
\begin{align}
&\alpha_Y(m_t)=0.01023,\quad  \alpha_2(m_t)=0.03337,\quad  \alpha_s(m_t)=0.1074 ~.  
\end{align}
The mass thresholds, discussed in Section~\ref{sec:betaf}, satisfy the following hierarchy: 
\begin{equation}
m_t<m_{\mathcal{L}}< \Lambda_{\rm PS} <m_{\mathcal P_6}<m_{\mathcal P_{10}}<\Lambda_{\mathrm{pre}} ~~. 
\end{equation}

\begin{figure}[t!]  
 \begin{center}
  \centering
  \hspace*{-0.3cm}
  \includegraphics[width=0.9\linewidth]{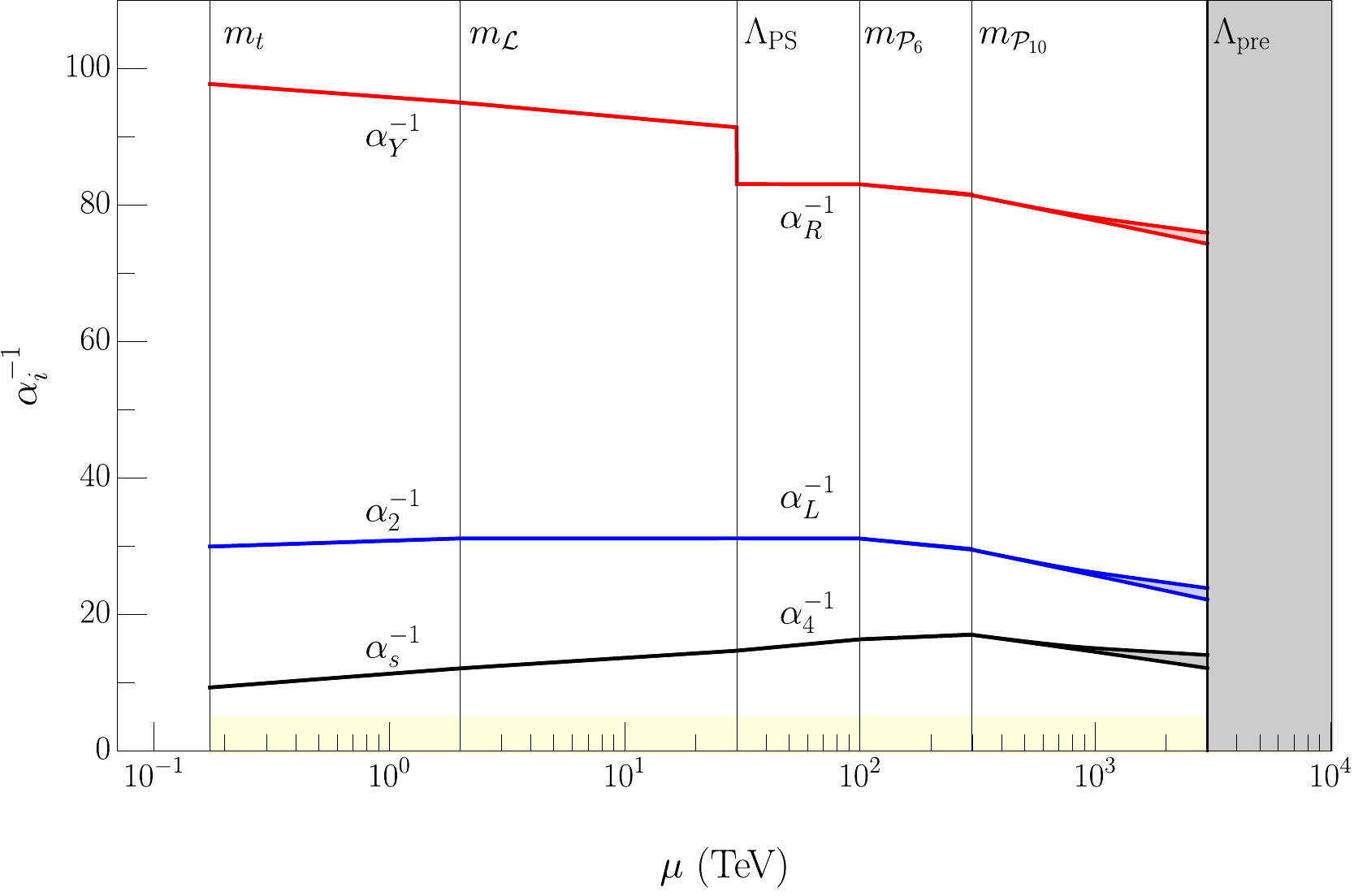}           
 \vspace*{0.1cm}
   \caption{Two-loop running of the inverse coupling constants ($\alpha^{-1}$) for 
   $SU(3)_{\rm c}  \times SU(2)_L \times U(1)_Y$ from the $m_t$ scale up to $\Lambda_{\rm PS}$, 
   and for     $SU(4)_{\rm PS} \times SU(2)_L \times SU(2)_R$ between $\Lambda_{\rm PS}$ and $\Lambda_{\rm pre}$. Vertical lines mark various thresholds.     
 Prebaryons $\Pb_{10}$,  $\Pb_{\overline{10}}$ and $\Pb_{15}$, assumed to have same mass $m_{\Pb_{10}}$, reduce the three $\alpha^{-1}$ 
  near $\Lambda_{\rm pre}$. Prebaryons $\Pb_{6,3}$ and $\Pb_{6,1}$ (of mass $m_{\Pb_6}$), as well as the charged di-prebaryons  and the composite vectorlike lepton 
   (of mass $m_{\cal L}$),  have a similar but less notable impact.  
   The 2-loop effects of the composite singlet scalars are shown in each lower line fork near $\Lambda_{\rm pre}$. 
  The gray-shaded vertical band at $\Lambda_{\rm pre}$ represents the region of strongly-coupled preonic $SU(15)_{\rm p}$ dynamics.
  The yellow-shaded horizontal band indicates the region where the three $\alpha$'s would no longer be perturbative.  
   \label{fig:running}
  }   
  \end{center}
\end{figure}

Using a reasonable set of values for these mass thresholds, we show the evolution of the inverse coupling constants in Figure~\ref{fig:running}. 
There, we set the mass of the composite vectorlike lepton at $m_{\mathcal{L}} = 2$ TeV (the current lower limit is 1 TeV, see Section~\ref{sec:belowPS}). 
For the scale of $SU(4)_{\text{PS}} \times SU(2)_R$ breaking we take $\Lambda_{\rm PS} =  30$ TeV. Note that in perturbative, minimal Pati-Salam models, constraints from rare meson decays impose a lower limit on the symmetry breaking scale of 80 TeV \cite{Dolan:2020doe,Valencia:1994cj}. 
However, in the presence of vectorlike fermions, that limit can be lowered to about 5 TeV \cite{Dolan:2020doe,Calibbi:2017qbu}.  

The remaining composite vectorlike fermions, which are heavier as they belong to higher representations of  $SU(4)_{\text{PS}} $ (see Table~\ref{table:PSprebaryons}), are assumed to have masses at $m_{\mathcal P_6} = 100$ TeV (for the $\Pb_{6,3}$ and $\Pb_{6,1}$ prebaryons) and $m_{\mathcal P_{10}} = 300$ TeV (for the $\Pb_{15}$, $\Pb_{10}$ and $\Pb_{\overline{10}}$ prebaryons). The benchmark choice for confinement scale $\Lambda_{\rm pre} = 3\times10^3$ TeV, is in line with the bounds from proton decay estimated in Ref.~\cite{Assi:2022jwg}. 

As can be seen in Figure~\ref{fig:running}, above $m_{\mathcal{L}}$ the three gauge couplings run to larger values compared to the SM values at the same scale. 
The strong coupling starts decreasing faster above $\Lambda_{\rm PS}$ due to the additional non-Abelian gauge bosons. The hypercharge coupling is replaced by the $SU(2)_R$ coupling at $\Lambda_{\rm PS}$, according to the matching condition (\ref{eqn:SMPS}). As expected based on Eqs.~(\ref{eq:abovePSb}) and (\ref{eqn:betaL2}), the $SU(2)_L \times SU(2)_R$ couplings remain almost constant at scales in the $\Lambda_{\rm PS} - m_{\mathcal P_6}$ range, while above the $m_{\mathcal P_6}$ scale they lose asymptotic freedom.
Above the scale of the heaviest chiral prebaryons, $m_{\mathcal P_{10}}$, the $SU(4)_{\rm PS}$ interaction also loses asymptotic freedom, but easily remains perturbative up to the compositeness scale $\Lambda_{\mathrm{pre}}$.
The two-loop effects of the gauge-singlet scalars are shown in Figure~\ref{fig:running} for each coupling as the lower branch near $\Lambda_{\rm pre}$, with 
the starting value (at the scalar masses) for all Yukawa couplings taken to be $y \approx 2$.  For comparison, the upper branches show the results without Yukawa couplings.

\subsection{Compositeness scale and premeson effects}
\label{scn:premesons}

Understanding the behavior of the $SU(4)_{\rm PS} \times SU(2)_L \times SU(2)_R$ gauge couplings near the compositeness scale \(\Lambda_{\mathrm{pre}}\) is 
challenging due to the presence of numerous charged fields and the nonperturbative effects of the $SU(15)_{\rm p}$ interactions.
Nevertheless, at least a rough estimate of the shifts in the gauge couplings near  $\Lambda_{\mathrm{pre}}$ is necessary for assesing the viability of the theory analyzed here. 
To see that, recall that we already encountered (below $\Lambda_{\mathrm{pre}}$) a loss of asymptotic freedom in the gauge couplings of $SU(2)_L \times SU(2)_R$ above the scale $m_{{\cal P}_6}$, and of  $SU(4)_{\rm PS}$ above the scale $m_{{\cal P}_{10}}$,
 due to the presence of several prebaryons charged under these gauge groups.
Moreover, above \(\Lambda_{\mathrm{pre}}\), the theory includes multiple preons that contribute to the running of the gauge couplings, preventing asymptotic freedom (see Section \ref{sec:dynamics}) and making it difficult for the couplings to remain perturbative towards higher energy scales. 

To investigate this, we consider the impact on the coupling running of the  premesons of spin-1 that are charged under $SU(4)_{\rm PS} \times SU(2)_L \times SU(2)_R$. 
These premesons have masses around the compositeness scale \(\Lambda_{\mathrm{pre}}\), and 
provide an uplift of the three inverse coupling constants. 
The specific premesons formed from the preon fields \(\Psi_W\), \(\Psi_W'\), and \(\psi_{i}\) are constructed as bilinears of the form 
\(\overline{\Psi}_{W} \sigma^\mu \Psi_{W'}\), \(\overline{\Psi}_{W} \sigma^\mu \Psi_{W}\), \(\overline{\psi}_i \sigma^\mu \Psi_{W}\), etc.
They are labelled by \(\prm_{_{WW'}} \, , \)  \(\prm_{_{WW}} \, ,  \)  \(\prm_{_{iW}} \, , \)  etc., and transform under the gauge representations listed in Table~\ref{table:PSmesons}. 
  Due to their large quadratic Casimir invariants or large multiplicities, 
these spin-1 fields  substantially decrease the $SU(4)_{\rm PS} \times SU(2)_L \times SU(2)_R$ gauge couplings.

\begin{table}[b!]
\begin{center}
\bigskip
\renewcommand{\arraystretch}{1.6}
\begin{tabular}{|c|c|c|}\hline    
\parbox[t]{1.9cm}{\ \ \ Spin-1 \\[-0.5mm] premesons \\[-3mm]  } & \parbox[t]{3.4cm}{ $\,$ \\[-3mm] \hspace*{0.3cm} Preon content  }    & 
\parbox[t]{5.6cm}{ 
\hspace*{0.1cm} $SU(4)_{\rm PS} \times SU(2)_L \times SU(2)_R$  \\[-0.1mm]  \hspace*{1cm}  representations }    \\ \hline \hline
$\prm_{_{WW'}}$ , $\prm_{_{W'W}}$  & $\overline{\Psi}_{W} \, \sigma^\mu \Psi_{W'}$ ,  $\overline{\Psi}_{W'} \, \sigma^\mu \Psi_{W}$    &  $(6,2,2) + (\overline{10},2,2) \; , \;  \text{conjugates}$ \\ \hline
$\prm_{_{WW}}$    & $\overline{\Psi}_{W} \, \sigma^\mu \Psi_{W}$     &  $(15,3,1) + (15,1,1) + (1,3,1)$ \\ \hline
$\prm_{_{W'W'}}$  &  $\overline{\Psi}_{W'} \, \sigma^\mu \Psi_{W'}$    &  $(15,1,3) + (15,1,1) + (1,1,3)$ \\ \hline 
$\prm_{_{i W}}$ , $\prm_{_{W i}}$   &  $\overline{\psi}_i \, \sigma^\mu \Psi_{W}$  , $\overline{\Psi}_W \, \sigma^\mu \Psi_i$  &  $3\times (4,2,1)  \; , \;  \text{conjugates}$ \\ \hline    
$\prm_{_{i W'}}$ , $\prm_{_{W' i}}$  &  $\overline{\psi}_i \, \sigma^\mu \Psi_{W'}$ , $\overline{\Psi}_{W'} \, \sigma^\mu \Psi_i$  &  $3\times (\overline{4},1,2)  \; , \;  \text{conjugates}$ \\ \hline 
\end{tabular}
\vspace*{2mm}
\caption{Spin-1 premesons (of mass around \(\Lambda_{\rm pre}\))   transforming under the $SU(4)_{\rm PS} \times SU(2)_L \times SU(2)_R$ gauge group. The generation index is \(i = 1, 2, 3\).}
\label{table:PSmesons}
\end{center}
\end{table}

The one-loop contributions from all spin-1 premesons on the coefficient of the beta function for an \(SU(N)\) gauge group ($N = 4$ or 2 in our case) is 
\bear
b(\rho) = \frac{11}{3} \left[ N_{\text{Ad}} \, C_2(\text{Ad}) + N_{S} \, C_2(S)+ N_{A} \, C_2(A)  + N_{F} \, C_2(F) \rule{0mm}{4mm}   \right] ~~.
\eear
Here \(N_{\text{Ad}}\), \(N_{S}\), \(N_{A}\),  and \(N_{F}\) are the numbers of spin-1 premesons transforming in the adjoint, symmetric, antisymmetric, and fundamental representations, as outlined in Table~\ref{table:PSmesons}. 
Note that $b(\rho) > 0 $ because the contribution of spin-1 fields in non-Abelian gauge theories is towards asymptotic freedom. 
In the specific case of \(SU(4)_{\rm PS}\), the quadratic Casimir invariants for these representations are \(C_2(\text{Ad}) = 4\), \(C_2(S) = 9/2 \), \(C_2(A) = 5/2 \),  and \(C_2(F) = 15/8 \). Using the number of pre-mesons in each representation (specifically, \(N_{\rm Ad} = N_A = N_S = 8\),  \(N_F = 24\)), the beta function coefficient for the \(SU(4)_{\rm PS}\) gauge coupling is found to be very large, \(b_4 (\rho)  = 1463/3  \). Similarly, \(b_{L,R}(\rho) = 1078/3 \), which is almost two orders of magnitude larger than typical one-loop beta function coefficients in the SM. 

Besides spin-1 premesons, there are also many spin-0 premesons and vectorlike prebaryons  
with masses around the  \(\Lambda_{\rm pre}\) scale. Their effect is harder to estimate, but it is in the opposite direction compared to the spin-1 states. Thus, the large positive contributions \(b_4 (\rho) \) and \(b_{L,R}(\rho) \) are at least partially compensated by other prehadrons. Both the sign and size of contributions from higher-spin prehadrons are difficult to determine. 
In addition, the RGEs of $\alpha_4$, $\alpha_L$, and $\alpha_R$, are affected by purely nonperturbative effects due to $SU(15)_{\rm p} $ dynamics, which presently cannot be  computed in chiral gauge theories. 

To obtain a rough lower bound for $\alpha_4$, $\alpha_L$, and $\alpha_R$ at a scale above preon confinement, we estimate the change in the gauge coupling due to the spin-1 premesons listed in Table~\ref{table:PSmesons}. The question is over what range of scales do the spin-1 premesons act as relevant degrees of freedom? For a rough estimate,  we use the information provided by QCD, although one should keep in mind that the behavior of spin-1 bound state in our chiral preon dynamics may be very different from QCD.
The measured mass of the \(\rho\) mesons is $m_\rho \approx 770$ MeV, while the scale where quarks and gluons become perturbative particles is around 2 GeV (see, e.g., \cite{ParticleDataGroup:2024cfk}). Thus, in QCD the \(\rho\) meson is a relevant particle up to a scale of about $2.6\, m_\rho$. 

Note that the running of the electromagnetic coupling constant $\alpha$ is pushed by \(\rho\) mesons towards larger values because QED is an Abelian gauge theory. Therefore,  we cannot use information about $\alpha$ to draw conclusions about $\alpha_4$, $\alpha_L$, and $\alpha_R$ (for a lattice study of QCD effects on $\alpha$, see \cite{Jegerlehner:2011mw}). 

For a preonic confinement scale of \(\Lambda_{\rm pre} \approx 3\times10^3\) TeV, consistent with proton decay constraints from Ref.~\cite{Assi:2022jwg}, the gauge coupling constants obtained in Section~\ref{sec:comb} are
\begin{equation}
\alpha_4(\Lambda_{\rm pre})=0.071,\quad
\alpha_L(\Lambda_{\rm pre})=0.042,\quad
\alpha_R(\Lambda_{\rm pre})=0.013 ~.
\end{equation}
If the influence of the \(\rho\) premesons persists up to a scale \( \Lambda'_{\rm pre} \approx 2.6 \Lambda_{\rm pre} \approx 8\times10^3 \) TeV (analogous to the endpoint of the \(\rho\) meson regime in QCD), then we find that the 1-loop running gives the following coupling ratios:
\begin{equation}
\frac{\alpha_4^{-1}(\Lambda'_{\rm pre})}{\alpha_4^{-1}(\Lambda_{\rm pre})}=6.1 ~,\quad
\frac{\alpha_L^{-1}(\Lambda'_{\rm pre})}{\alpha_L^{-1}(\Lambda_{\rm pre})}=3.2 ~,\quad
\frac{\alpha_R^{-1}(\Lambda'_{\rm pre})}{\alpha_R^{-1}(\Lambda_{\rm pre})}=1.7 ~.
\end{equation}
These large ratios illustrate the substantial impact that the premesons have on the running of gauge couplings.  
Thus, $SU(15)_{\rm p} $ dynamics may substantially push up the three $\alpha^{-1}$ running curves shown in Figure~\ref{fig:running} once they enter the right-handed (gray) shaded region. 

At scales beyond $\Lambda_{\rm pre}$, where the preons rather than the prehadrons are the relevant degrees of freedom,  the 
$SU(4)_{\rm PS} \times SU(2)_L \times SU(2)_R$ interactions lose again asymptotic freedom (see discussion in  Section \ref{sec:dynamics}). However, the decrease of the couplings due to  $SU(15)_{\rm p} $ dynamics around $\Lambda_{\rm pre}$ may be sufficient to keep $\alpha_4$, $\alpha_L$, and $\alpha_R$ perturbative up to a unification scale $\Lambda_{422}$.   
There,  $SU(4)_{\rm PS} \times SU(2)_L \times SU(2)_R$ may be embedded in the \(SO(10)\) gauge group, which is large enough to have asymptotic freedom in the presence of all the preons. 

Whether the three couplings unify at the $\Lambda_{422}$ scale depends on various threshold effects and symmetry breaking patterns \cite{Giunti:1991ta}. Compared to usual, perturbative GUT models, the unification scale can be higher ({\it e.g.}, $\Lambda_{422} \approx 10^{17}$ GeV) in the framework discussed here due to the intricate running of the three gauge couplings. As a result, Planck-scale suppressed operators may be large enough to already split  $\alpha_4$, $\alpha_L$, and $\alpha_R$ at $\Lambda_{422}$.

\section{Conclusions}
\label{sec:conclusion}

We have developed a comprehensive framework for quark and lepton compositeness, based on an \(SU(15)_{\rm p}\) chiral gauge theory that confines preons. This theory not only yields three SM generations of composite quarks and leptons (as 3-preon bound states), but also provides a dynamical origin for symmetry breaking and the Higgs sector (as 6-preon bound states). The preons are also charged under the weakly-coupled \(SU(4)_{\rm PS} \times SU(2)_L \times SU(2)_R\) gauge group, which is broken down to the SM gauge group at a scale $\Lambda_{\rm PS}$ in the $30-100$ TeV range. 

By embedding the QCD gauge group within \(SU(4)_{\rm PS}\),  the ultraviolet behavior of the strong coupling is moderated, at least up to the preon confinement scale $\Lambda_{\rm pre}$, which is about two orders of magnitude above $\Lambda_{\rm PS}$. 
Our analysis of renormalization group evolution, incorporating the effects of composite vectorlike fermions and scalar di-prebaryons, shows that each of the \(SU(4)_{\rm PS} \times SU(2)_L \times SU(2)_R\) interactions lose asymptotic freedom below $\Lambda_{\rm pre}$. Near the compositeness scale there are many spin-1 premesons whose effect is to decrease the gauge couplings. If that effect is not fully counterbalanced by other prehadrons and nonperturbative effects at $\Lambda_{\rm pre}$ that cannot be currently computed, the   \(SU(4)_{\rm PS} \times SU(2)_L \times SU(2)_R\) gauge group may remain under perturbative control up to a unification scale, where it may be embedded in \(SO(10)\).

The theory opens avenues for novel mechanisms of quark and lepton mass generation rooted in underlying preon dynamics. 
It also predicts composite vectorlike leptons not far above the TeV scale, as well as an extended Higgs sector, offering observable signatures at the LHC and future colliders. We have found that the vectorlike fermion spectrum is modified in the presence of the  \(SU(4)_{\rm PS} \times SU(2)_L \times SU(2)_R\) gauge interactions, so the collider probes of this theory are different compared to the case where the weakly-coupled gauge symmetry is just the SM one \cite{Dobrescu:2021fny, Assi:2022jwg}.
Moreover, the composite neutrino sector is complex, with several sterile neutrinos that may have interesting phenomenological implications.

There remain, however, many important questions, especially related to the behavior of the strongly-coupled chiral \(SU(15)_{\rm p}\) gauge interactions. For example, the size of the VEVs of di-prebaryons is highly dependent on the coupling strength in the attractive channels.  
Consequently, a quantitative estimate of the mass spectrum for the composite vectorlike fermions  is 
hampered by the lack of information regarding how close to the critical value is the binding due to premeson exchange. 

Furthermore, the possibility that the preonic bound states associated with the Higgs sector have masses several orders of magnitude below \(\Lambda_{\rm pre}\) needs to be put on a firmer ground. 
Nevertheless, the framework presented here offers a promising path for the preonic  \(SU(15)_{\rm p}\) chiral gauge dynamics to be a realistic substructure origin of the SM.

\bigskip\bigskip\bigskip\bigskip

{\bf Acknowledgments:} \ We would like to thank Jure Zupan and Ryan Plestid for helpful conversations. B.A. acknowledges support in part by the DOE grant DE-SC1019775, and the NSF grants OAC-2103889, OAC-2411215, and OAC- 2417682.  
B.A.'s  work was performed in part at the Aspen Center for Physics, with support by a grant from the Simons Foundation (1161654,Troyer).
B.D.'s work was supported by Fermi Forward Discovery Group, LLC under Contract No. 89243024CSC000002 with the U.S. Department of Energy, Office of Science, Office of High Energy Physics.

\smallskip


\end{document}